\documentclass[11pt,a4paper]{article}
\pdfoutput=1

\usepackage{jheppub}

\usepackage{amsmath}
\usepackage{bbm}
\usepackage{float}
\usepackage{graphicx}	
\usepackage{hyperref}
\usepackage{mathtools}
\usepackage{cancel}

\usepackage[normalem]{ulem}

\usepackage{multirow}
\usepackage{rotating}
\usepackage{subcaption}

\def\lsim{\raise0.3ex\hbox{$\;<$\kern-0.75em\raise-1.1ex
\hbox{$\sim\;$}}}
\def\gsim{\raise0.3ex\hbox{$\;>$\kern-0.75em\raise-1.1ex
\hbox{$\sim\;$}}}

\def\thetitle{ 
Toward diagnosing neutrino non-unitarity through CP phase correlations \\
 \vspace{- 6mm}
}
\title{\thetitle}
\hypersetup{pdftitle={\thetitle}}
\author{Hisakazu Minakata}
\affiliation{
Center for Neutrino Physics, Department of Physics, Virginia Tech, Blacksburg, Virginia 24061, USA \\
}

\emailAdd{hisakazu.minakata@gmail.com}
\date{\today}

\abstract{ 
We discuss correlations between the $\nu$SM CP phase $\delta$ and the phases that originate from new physics which causes neutrino-sector unitarity violation (UV) at low energies. This study is motivated to provide one of the building pieces for a machinery to diagnose non-unitarity, our ultimate goal. We extend the perturbation theory of neutrino oscillation in matter proposed by Denton {\it et al.}~(DMP) to include the UV effect expressed by the $\alpha$ parametrization. By analyzing the DMP-UV perturbation theory to first order, we are able to draw a completed picture of the $\delta$ - UV phase correlations in the whole kinematical region covered by the terrestrial neutrino experiments. There exist the two regions with the characteristically different patterns of the correlations: 
(1) the chiral-type $[e^{- i \delta } \alpha_{\mu e}, ~e^{ - i \delta} \alpha_{\tau e}, ~\alpha_{\tau \mu}]$ (PDG convention) correlation in the entire high-energy region $\vert \rho E \vert \gsim 6~(\text{g/cm}^3)$ GeV, and (2) (blobs of the $\alpha$ parameters) - $e^{ \pm i \delta}$ correlation in anywhere else. Some relevant aspects for measurement of the UV parameters, such as the necessity of determining all the $\alpha_{\beta \gamma}$ elements at once, are also pointed out. 

} 

\newcommand{\Dmsqren}{\Delta m^2_{ \text{ren} }}

\newcommand{\ket}[1]{|#1\rangle}

\begin{document} 

\maketitle

\section{Introduction}
\label{sec:introduction}

At more than forty years after the establishment of the Standard Model (SM) of electroweak interactions~\cite{Weinberg:1967tq,Glashow:1961tr,Salam:1968rm} in the seventies, people naturally sought discovery for physics beyond the SM. Despite we now know that neutrinos are massive and the lepton flavors mix~\cite{Kajita:2016cak,McDonald:2016ixn}, and we have the compelling evidences for dark matter in the universe~\cite{Bertone:2004pz,Villanueva-Domingo:2021spv}, a conclusive bigger picture of our fundamental world does not appear to be born. Naturally one asks the question: If we interpret the above open windows as suggestions for the right places to look for new physics beyond the SM, what should we do? 

In neutrino physics a possibility of existence of SM-singlet states, or sterile neutrinos, is widely discussed, as reviewed e.g., in refs.~\cite{Dasgupta:2021ies,Dentler:2018sju,Conrad:2013mka} and the references cited therein. A version of it, sterile leptons with eV scale masses might have already been seen in the LSND~\cite{LSND:2001aii} and MiniBooNE~\cite{MiniBooNE:2018esg} experiments. However, the very recent MicroBooNE data seem to disfavor both the $\nu_e$~\cite{MicroBooNE:2021rmx} and photon~\cite{MicroBooNE:2021zai} origins of the low energy excess of MiniBooNE.\footnote{
For a different interpretation of the same data, see ref.~\cite{Denton:2021czb}.
}
On the other hand, IceCube sees a closed contour at 90\% CL in the mixing angle - $\Delta m^2$ space, which is interpreted as a ``systematic effect'', not a fluke~\cite{IceCube:2020tka}. Thus, the chance of settling the tantalizing question of eV scale sterile neutrino(s) {\em yes or no} is left for the ongoing and upcoming searches, which is to be joined by those in refs.~\cite{Machado:2019oxb,JSNS2:2021hyk}. 
In more generic contexts, search for deviation from the SM expectation is done in the frameworks of so called the ``non-standard interactions'' (NSI)~\cite{Wolfenstein:1977ue}, and/or non-unitarity~\cite{Antusch:2006vwa,Escrihuela:2015wra}.\footnote{
Given the feature of generic NSI with $9 \times 3=27$ parameters, non-unitarity in matter can be regarded as a ``constrained NSI'' with 9 parameters only, in which the production and detection NSI elements are determined by the propagation NSI. See the discussions in refs.~\cite{Blennow:2016jkn,Martinez-Soler:2018lcy,Fong:2017gke}. 
}
See e.g. refs.~\cite{Ohlsson:2012kf,Miranda:2015dra,Farzan:2017xzy,Proceedings:2019qno} for reviews of NSI, refs.~\cite{Antusch:2008tz,Biggio:2009nt,Esteban:2018ppq} for constraints on NSI, and refs.~\cite{Antusch:2006vwa,Escrihuela:2015wra,Blennow:2016jkn,Fong:2016yyh,Fong:2017gke,Martinez-Soler:2018lcy,Martinez-Soler:2019noy,Fernandez-Martinez:2007iaa,Goswami:2008mi,Antusch:2009pm,Antusch:2009gn,Antusch:2014woa,Ge:2016xya,Fernandez-Martinez:2016lgt,Dutta:2016vcc,Parke:2015goa,Ellis:2020hus,Coloma:2021uhq} for a limited list of articles on non-unitarity. 

In this paper we address the non-unitarity approach. Let us imagine under which scenery the non-unitarity will be studied. If the eV scale sterile neutrino is the cause of non-unitarity, it is likely that its presence and the properties will be known by the advanced searches such as in refs.~\cite{IceCube:2020tka,Machado:2019oxb,JSNS2:2021hyk} in the very near future, unless their mixing to the active sector is extremely small. With positive evidence for accessible low-mass sterile(s), one can just go to the experimental data to dig out the correct shape of the sterile lepton model. 

Suppose, however, that it does not happen, but nonetheless the results of various precision measurements continue to report small but robust deviation from the $\nu$SM, a shorthand notation for the neutrino-mass embedded SM. Then, the obvious question must be: What is detected? A possible suspect would be non-unitarity in more generic sense. It may be described by relatively model-independent frameworks, such as the one for physics at high scale~\cite{Antusch:2006vwa}, $E \gg m_{W}$, or at low scale~\cite{Fong:2016yyh,Fong:2017gke}, $E \ll m_{W}$. If we lack the obvious candidates for such anomaly, we need to identify the nature of physics behind non-unitarity either by the phenomenological methods, or preferably via the experimental ways. 

In this paper, we investigate correlations between the $\nu$SM CP phase and the phases originated from new physics existing at some scale, which is recognized as unitarity violation (UV)\footnote{
We are aware that in physics literatures UV usually means ``ultraviolet''. But, in this paper UV is used as an abbreviation for ``unitarity violation'' or ``unitarity violating''.  }
at low energies. With use of the UV-extended version of Denton {\it et al.} framework~\cite{Denton:2016wmg}, we aim at establishing a unified view of such correlations valid in the whole kinematical region covered by the terrestrial neutrino experiments, the ``terrestrial region'', for short. Typically this is the region of Super-Kamiokande's observation of atmospheric neutrinos, 0.1 GeV $\lsim E \lsim 10$ GeV, as reported in Fig.~3 in ref.~\cite{Super-Kamiokande:2017yvm}. A rough sketch of the equi-probability contour of $P(\nu_{\mu} \rightarrow \nu_{e})$ in this region is drawn in Fig.~1 of ref.~\cite{Minakata:2019gyw}. Our treatment of the phase correlations in this paper surpasses those of the previous works~\cite{Martinez-Soler:2018lcy,Martinez-Soler:2019noy} which apply only to the two local regions of atmospheric- and solar-scale enhanced oscillations, roughly speaking, the resonances~\cite{Wolfenstein:1977ue,Barger:1980tf,Mikheyev:1985zog,Smirnov:2016xzf}.

The ultimate goal in our approach is to diagnose non-unitarity which may be originated in new physics beyond the $\nu$SM. It is conceivable that the effect of non-unitarity starts to be seen in the interference term $S_{\nu\text{SM}}^{*} S_{ \text{UV} }$, where $S_{ \text{UV} }$ denotes the UV amplitude. See Appendix~\ref{sec:DMP-UV-formulation} for the more concrete form of $S_{ \text{UV} }$. In this approach, what should be reached at the end is to extract the physical properties of $S_{ \text{UV} }$ by analyzing the experimental data which include its effect in the form of $S_{\nu\text{SM}}^{*} S_{ \text{UV} }$, {\it diagnosing non-unitarity}. 

Toward the goal, we attempt at a theoretical investigation of the $\nu$SM $e^{ \pm i \delta}$ - UV phase correlation involved in the interference term $S_{\nu\text{SM}}^{*} S_{ \text{UV} }$ in this paper. Notice that $e^{ \pm i \delta}$ is the unique phase factor in the $\nu$SM oscillation amplitude which is written by the fundamental physical parameter of the $\nu$SM, the lepton KM phase $\delta$~\cite{Kobayashi:1973fv}. Therefore, its correlation with UV phase factor must contain the key information on the relationship between $\nu$SM and UV new physics. We will return to the issue of possible relevance of our investigation of $\nu$SM - UV phase correlations in the entire program of diagnosing non-unitarity in sections~\ref{sec:completed-picture} and \ref{sec:conclusion}. 

This strategy of diagnosing non-unitarity through interference between the $\nu$SM and UV-driving new physics presumes that such correlation exists at a detectable level even in the case that new physics scale is much higher than $m_{W}$. Though it may sound unlikely, it is not totally obvious if we can conclude it impossible. 
We all know that physics must exist at much higher energy scale as $1 / \sqrt{G_N} \sim 10^{19}$ GeV, but almost massless particles around us are allowed to exist. As far as the formulation of high-scale UV is correct in deriving the non-unitary flavor mixing matrix, the $\nu$SM and UV-driving new physics should interfere.\footnote{
If this or the possible other reasonings fail, low-scale UV might be a more natural scenario to expect the $\nu$SM-UV correlations. }
We also add that if CP violation has a group-theoretical origin related to strings~\cite{Ratz:2019zak}, the correlations between the phases should exist, carrying crucially important information in such system. 

Despite that a slight overlap may exist between our descriptions of this paper and the ones in refs.~\cite{Martinez-Soler:2018lcy,Martinez-Soler:2019noy}, we will try to make this paper self-contained as much as possible. 

\section{$\nu$SM - UV phase correlations: Now and the next step}
\label{sec:now-and-next} 

To discuss the $\nu$SM - UV phase correlations in an unambiguous way, we must first decide the way how the UV effect is parametrized. We use so called the $\alpha$ parametrization~\cite{Escrihuela:2015wra} in which the non-unitary flavor mixing matrix $N$ is defined by multiplication of the $\alpha$ matrix to the usual unitary $\nu$SM mixing matrix $U \equiv U_{\text{\tiny MNS}}$~\cite{Maki:1962mu} in the Particle Data Group (PDG) convention~\cite{Zyla:2020zbs}~(see eq.~\eqref{MNS-PDG}) as 
\begin{eqnarray} 
N_{\text{\tiny PDG}}
&=& 
\left( \bf{1} - \alpha \right) U_{\text{\tiny PDG}}
= 
\left\{ 
\bf{1} - 
\left[ 
\begin{array}{ccc}
\alpha_{ee} & 0 & 0 \\
\alpha_{\mu e} & \alpha_{\mu \mu}  & 0 \\
\alpha_{\tau e}  & \alpha_{\tau \mu} & \alpha_{\tau \tau} \\
\end{array}
\right] 
\right\}
U_{\text{\tiny PDG}}. 
\label{alpha-matrix}
\end{eqnarray}
The $\alpha$ parametrization originates in refs.~\cite{Okubo:1961jc,Schechter:1980gr}. The $\alpha$ matrix has nine degrees of freedom due to the three real diagonal and the three complex off-diagonal entries. 

The problem of correlation between the CP phase factor $e^{ \pm i \delta}$ in the $\nu$SM  and the one in the UV amplitude is investigated in previous papers~\cite{Martinez-Soler:2018lcy,Martinez-Soler:2019noy} using the $\alpha$ parametrization. In these references use has been made of the UV-extended frameworks of the ones given in refs.~\cite{Minakata:2015gra,Martinez-Soler:2019nhb}, whose former (latter) is for region around the atmospheric (solar) resonance. 
Interestingly, very different types of the phase correlation are observed in these two regions. A charming ``chiral'' type correlation $[e^{- i \delta } \alpha_{\mu e}, ~e^{ - i \delta} \alpha_{\tau e}, ~\alpha_{\tau \mu}]$ (PDG convention)  is found in the former~\cite{Martinez-Soler:2018lcy}, whereas in the latter a less transparent (blobs of the $\alpha$ parameters) - $e^{ \pm i \delta}$ correlation~\cite{Martinez-Soler:2019noy} is seen. It is good to know that a part of the chiral-type correlation, $e^{ - i \delta } \alpha_{\mu e}$, in the atmospheric region has been observed in the foregoing analyses~\cite{Escrihuela:2015wra,Miranda:2016wdr,Abe:2017jit}. 

\subsection{Toward $\nu$SM - UV phase correlations in the whole terrestrial region} 
\label{sec:terrestrial-view} 

Thus, to our current knowledge, the picture of $\nu$SM - UV phase correlation jumps from a local region to another, from the chiral-type correlation in the atmospheric resonance region to the $\delta - \alpha_{\beta \gamma}$-blobs correlation in the solar resonance region. Obviously, we need a better treatment of the phase correlations to allow us to draw a global picture of the phase correlation in the whole terrestrial region, i.e., the region covered by the terrestrial neutrino experiments. 

To our knowledge the right theoretical framework for this purpose is the one based on the Jacobi method, Denton {\it et al.} (DMP) perturbation theory~\cite{Denton:2016wmg} and Agarwalla {\it et al.} (AKT) perturbation theory~\cite{Agarwalla:2013tza} with full coverage of the whole terrestrial region. For a pedagogical introduction for the Jacobi method, see ref.~\cite{Agarwalla:2013tza}.\footnote{
It is shown that these Jacobi-method-based approximation schemes provide numerically accurate probability expressions~\cite{Parke:2019vbs}. For further studies of the Jacobi method in neutrino oscillation, see refs.~\cite{Denton:2018fex,Denton:2019qzn}. 
}
See Appendix~A in ref.~\cite{Minakata:2021goi} for another favorable feature of the globally valid frameworks, albeit it being a much less familiar one. 

We base our formalism on the DMP perturbation theory and extend it to incorporate the UV effect parametrized by the $\alpha$ parameters, the framework dubbed hereafter as the ``DMP-UV perturbation theory''. It is because DMP is easier to handle, and it has the transparent relation to a particular version~\cite{Minakata:2015gra} of the atmospheric-resonance perturbation theory as ``half a way'' to  DMP. In fact, it is shown analytically~\cite{Minakata:2020oxb} that the DMP theory approaches to this atmospheric-resonance perturbation theory and to the solar-resonance perturbation theory~\cite{Martinez-Soler:2019nhb}, respectively, each in the appropriate limit. In this way, we can discuss the relationship between our results in this paper and the ones obtained previously by using the atmospheric- and solar-resonance perturbation theories valid in these local regions~\cite{Martinez-Soler:2018lcy,Martinez-Soler:2019noy}. 

\subsection{Non-unitarity vs. sterile neutrino}
\label{sec:UV-vs-sterile} 

It may be worth to make clarifying remarks about what we mean by non-unitarity approach and its relation to sterile neutrino hypothesis, the both mentioned in Introduction. Existence of sterile neutrinos is a perfect model for non-unitarity, and their separate treatment merely reflects our preferred analysis strategy for physics beyond the $\nu$SM. Of course, the systems with non-unitarity have a much wider variety, as first raised in ref.~\cite{Antusch:2006vwa} and further studied in many references cited in section~\ref{sec:introduction}. 

Sterile neutrino states with eV scale masses can be searched for by the LSND-MiniBooNE type experiments, for example. See e.g., ref.~\cite{Dasgupta:2021ies} for the other options. Alternatively, it produces the resonance-like enhancement in $P(\nu_{\mu} \rightarrow \nu_{\mu})$ and $P(\nu_{\mu} \rightarrow \nu_{e})$ at energy $E \sim 1$ TeV~\cite{Yasuda:2000xs,Nunokawa:2003ep}. As examined in details by many authors, see the list e.g., in ref. \cite{Esmaili:2018qzu}, it would allow detection in atmospheric neutrino observations as is being pursuit by IceCube~\cite{IceCube:2020tka}. Once a (or a few) sterile neutrino state(s) is identified, ``diagnosing non-unitarity'' is no more necessary. One can just examine the data to create the model of sterile state(s). 

Thus, our non-unitarity approach when applied to sterile neutrinos, is meant to treat cases with more elusive sterile states. Let us take the three-active plus $N_{s}$ sterile neutrino model with sterile masses of eV to MeV to make a more concrete statements. We have argued that such model can provide a generic model for low-scale UV with the $\nu$SM-like three-active neutrino oscillation probability modified by the non-unitary mixing matrix~\cite{Fong:2017gke,Fong:2016yyh}. The condition by which we remain in such ``diagnostics needed'' regime is worked out to be $\vert \rho E \vert \lsim 100 (\text{g/cm}^3)$ GeV, see section 3.5 in ref.~\cite{Fong:2017gke}. At energies higher than this by a factor of $\sim$ 50 we meet the $\mathcal{O}(1)$ TeV resonance, and the model moves into the ``no need for diagnostics'' regime with this unmistakable signature. Therefore, while we aim validity of our discussion in this paper in the ``terrestrial region'', in fact it can extends to $\vert \rho E \vert \lsim 100 (\text{g/cm}^3)$ GeV. 



\section{Three active neutrino system with non-unitary flavor mixing matrix}
\label{sec:3nu-non-unitarity} 

To formulate the DMP-UV perturbation theory with use of the $\alpha$ parametrization, we follow the method developed in ref.~\cite{Martinez-Soler:2018lcy}. We define the system below but defer presentation of a step-by-step formulation of the DMP-UV perturbation theory to Appendix~\ref{sec:DMP-UV-formulation} where it will be done in a pedagogical manner. 

In studies for formulating the three active neutrino evolution in matter in the presence of non-unitary flavor mixing, they appear to have converged to a framework that starts from the Schr\"odinger equation in the vacuum mass eigenstate basis 
\begin{eqnarray}
i \frac{d}{dx} \check{\nu} = 
\frac{1}{2E} 
\left\{  
\left[
\begin{array}{ccc}
0 & 0 & 0 \\
0 & \Delta m^2_{21} & 0 \\
0 & 0 & \Delta m^2_{31} \\
\end{array}
\right] + 
N^{\dagger} \left[
\begin{array}{ccc}
a - b & 0 & 0 \\
0 & -b & 0 \\
0 & 0 & -b \\
\end{array}
\right] N 
\right\} 
\check{\nu}. 
\label{evolution-check-basis}
\end{eqnarray}
We just quote refs.~\cite{Blennow:2016jkn} and \cite{Fong:2017gke} for high-scale and low-scale UV, respectively, to support our statement. In the latter it is a truncated system from the 3 active + $N_{s}$ sterile model. In this paper, we denote the vacuum mass eigenstate basis as the ``check basis''. In eq.~(\ref{evolution-check-basis}), $N$ denotes the $3 \times 3$ non-unitary flavor mixing matrix which relates the flavor neutrino states to the vacuum mass eigenstates as 
\begin{eqnarray}
\nu_{\alpha} = N_{\alpha i} \check{\nu}_{i}. 
\label{N-def}
\end{eqnarray}
Hereafter, the subscript Greek indices $\alpha$, $\beta$, or $\gamma$ run over $e, \mu, \tau$, and the Latin indices $i$, $j$ run over the mass eigenstate indices $1,2,$ and $3$. $E$ is neutrino energy and $\Delta m^2_{ji} \equiv m^2_{j} - m^2_{i}$. The usual phase redefinition of neutrino wave function is done to leave only the mass squared differences.

The functions $a(x)$ and $b(x)$ in eq.~(\ref{evolution-check-basis}) denote the
Wolfenstein matter potential \cite{Wolfenstein:1977ue} due to charged current (CC) and neutral current (NC) reactions, respectively.
\begin{eqnarray} 
a(x) &=&  
2 \sqrt{2} G_F N_e E \approx 1.52 \times 10^{-4} \left( \frac{Y_e \rho}{\rm g\,cm^{-3}} \right) \left( \frac{E}{\rm GeV} \right) {\rm eV}^2, 
\nonumber \\
b(x) &=& \sqrt{2} G_F N_n E = \frac{1}{2} \left( \frac{N_n}{N_e} \right) a, 
\label{matt-potential}
\end{eqnarray}
where $G_F$ is the Fermi constant. 
$N_e$ and $N_n$ are the electron and neutron number densities in matter. $\rho$ and $Y_e$ denote, respectively, the matter density and number of electrons per nucleon in matter. These four quantities are, in principle, position dependent. 

\subsection{The three useful conventions of the lepton flavor mixing matrix}
\label{sec:3-conventions}

We start from the most commonly used form, the PDG convention~\cite{Zyla:2020zbs} of the MNS matrix, 
\begin{eqnarray} 
U_{\text{\tiny PDG}} = 
\left[
\begin{array}{ccc}
1 & 0 &  0  \\
0 & c_{23} & s_{23} \\
0 & - s_{23} & c_{23} \\
\end{array}
\right] 
\left[
\begin{array}{ccc}
c_{13}  & 0 & s_{13} e^{- i \delta} \\
0 & 1 & 0 \\
- s_{13} e^{ i \delta} & 0 & c_{13}  \\
\end{array}
\right] 
\left[
\begin{array}{ccc}
c_{12} & s_{12}  &  0  \\
- s_{12} & c_{12} & 0 \\
0 & 0 & 1 \\
\end{array}
\right].
\label{MNS-PDG}
\end{eqnarray}
with the obvious notations $s_{ij} \equiv \sin \theta_{ij}$ etc. and $\delta$ being the CP violating phase. Recently, we have started to use the other two conventions different only by phase redefinitions called the ATM and SOL conventions: 
\begin{eqnarray} 
&&
U_{\text{\tiny ATM}} 
\equiv 
\left[
\begin{array}{ccc}
1 & 0 &  0  \\
0 & 1 & 0 \\
0 & 0 & e^{ - i \delta} \\
\end{array}
\right] 
U_{\text{\tiny PDG}} 
\left[
\begin{array}{ccc}
1 & 0 &  0  \\
0 & 1 & 0 \\
0 & 0 & e^{ i \delta} \\
\end{array}
\right] 
=
\left[
\begin{array}{ccc}
1 & 0 &  0  \\
0 & c_{23} & s_{23} e^{ i \delta} \\
0 & - s_{23} e^{- i \delta} & c_{23} \\
\end{array}
\right] 
\left[
\begin{array}{ccc}
c_{13}  & 0 & s_{13}  \\
0 & 1 & 0 \\
- s_{13} & 0 & c_{13} \\
\end{array}
\right] 
\left[
\begin{array}{ccc}
c_{12} & s_{12}  &  0  \\
- s_{12} & c_{12} & 0 \\
0 & 0 & 1 \\
\end{array}
\right], 
\nonumber \\
&&
U_{\text{\tiny SOL}} 
\equiv 
\left[
\begin{array}{ccc}
1 & 0 &  0  \\
0 & e^{ - i \delta} & 0 \\
0 & 0 & e^{ - i \delta} \\
\end{array}
\right] 
U_{\text{\tiny PDG}} 
\left[
\begin{array}{ccc}
1 & 0 &  0  \\
0 & e^{ i \delta} & 0 \\
0 & 0 & e^{ i \delta} \\
\end{array}
\right] 
= 
\left[
\begin{array}{ccc}
1 & 0 &  0  \\
0 & c_{23} & s_{23} \\
0 & - s_{23} & c_{23} \\
\end{array}
\right] 
\left[
\begin{array}{ccc}
c_{13}  & 0 & s_{13} \\
0 & 1 & 0 \\
- s_{13} & 0 & c_{13} \\
\end{array}
\right] 
\left[
\begin{array}{ccc}
c_{12} & s_{12} e^{ i \delta}  &  0  \\
- s_{12} e^{- i \delta} & c_{12} & 0 \\
0 & 0 & 1 \\
\end{array}
\right]. 
\nonumber \\
\label{ATM-SOL-def} 
\end{eqnarray}
The reason for our terminology of $U_{\text{\tiny ATM}}$ and $U_{\text{\tiny SOL}}$ in (\ref{ATM-SOL-def}) is because the CP phase factor $e^{ \pm i \delta}$ is attached to the ``atmospheric angle'' $s_{23}$ in $U_{\text{\tiny ATM}}$, and to the ``solar angle'' $s_{12}$ in $U_{\text{\tiny SOL}}$, respectively. Whereas in $U_{\text{\tiny PDG}}$, $\delta$ is attached to $s_{13}$. 

Once the phase convention of the $U$ matrix is changed from $U_{\text{\tiny PDG}}$ to $U_{\text{\tiny ATM}}$, a consistent definition of $N_{\text{\tiny ATM}}$ requires the $\alpha$ matrix to transform as can be seen in 
\begin{eqnarray} 
N_{\text{\tiny ATM}} 
&\equiv& 
\left[
\begin{array}{ccc}
1 & 0 &  0  \\
0 & 1 & 0 \\
0 & 0 & e^{ - i \delta} \\
\end{array}
\right] 
N_{\text{\tiny PDG}} 
\left[
\begin{array}{ccc}
1 & 0 &  0  \\
0 & 1 & 0 \\
0 & 0 & e^{ i \delta} \\
\end{array}
\right] 
= \left\{ 
\bf{1} - 
\left[
\begin{array}{ccc}
1 & 0 &  0  \\
0 & 1 & 0 \\
0 & 0 & e^{ - i \delta} \\
\end{array}
\right] 
\alpha 
\left[
\begin{array}{ccc}
1 & 0 &  0  \\
0 & 1 & 0 \\
0 & 0 & e^{ i \delta} \\
\end{array}
\right] 
\right\}
U_{\text{\tiny ATM}} 
\nonumber \\
&\equiv& 
\left( \bf{1} - \alpha^{\text{\tiny ATM}}  \right) U_{\text{\tiny ATM}}, 
\label{alpha-ATM-SOL}
\end{eqnarray}
and therefore the $\alpha$ matrix is convention dependent. It takes the form in the ATM convention 
\begin{eqnarray} 
&&
\hspace{-10mm}
\alpha^{\text{\tiny ATM}} 
= 
\left[
\begin{array}{ccc}
1 & 0 &  0  \\
0 & 1 & 0 \\
0 & 0 & e^{ - i \delta} \\
\end{array}
\right] 
\alpha 
\left[
\begin{array}{ccc}
1 & 0 &  0  \\
0 & 1 & 0 \\
0 & 0 & e^{ i \delta} \\
\end{array}
\right] 
=
\left[ 
\begin{array}{ccc}
\alpha_{ee} & 0 & 0 \\
\alpha_{\mu e} & \alpha_{\mu \mu}  & 0 \\
e^{ - i \delta} \alpha_{\tau e} & e^{ - i \delta} \alpha_{\tau \mu} & \alpha_{\tau \tau} \\
\end{array}
\right]
\equiv 
\left[ 
\begin{array}{ccc}
\alpha_{ee}^{\text{\tiny ATM}} & 0 & 0 \\
\alpha_{\mu e}^{\text{\tiny ATM}} & \alpha_{\mu \mu}^{\text{\tiny ATM}} & 0 \\
\alpha_{\tau e}^{\text{\tiny ATM}} & \alpha_{\tau \mu}^{\text{\tiny ATM}} & \alpha_{\tau \tau}^{\text{\tiny ATM}} \\
\end{array}
\right].
\label{alpha-ATM-def}
\end{eqnarray}
Similarly, $N_{\text{\tiny SOL}} \equiv \left( \bf{1} - \alpha^{\text{\tiny SOL}}  \right) U_{\text{\tiny SOL}}$ with 
\begin{eqnarray} 
&&
\alpha^{\text{\tiny SOL}} 
= 
\left[
\begin{array}{ccc}
1 & 0 &  0  \\
0 & e^{ - i \delta} & 0 \\
0 & 0 & e^{ - i \delta} \\
\end{array}
\right] 
\alpha 
\left[
\begin{array}{ccc}
1 & 0 &  0  \\
0 & e^{ i \delta} & 0 \\
0 & 0 & e^{ i \delta} \\
\end{array}
\right] 
=
\left[ 
\begin{array}{ccc}
\alpha_{ee} & 0 & 0 \\
e^{ - i \delta} \alpha_{\mu e} & \alpha_{\mu \mu}  & 0 \\
e^{ - i \delta} \alpha_{\tau e}  & \alpha_{\tau \mu} & \alpha_{\tau \tau} \\
\end{array}
\right] 
\equiv
\left[ 
\begin{array}{ccc}
\widetilde{\alpha}_{ee} & 0 & 0 \\
\widetilde{\alpha}_{\mu e} & \widetilde{\alpha}_{\mu \mu}  & 0 \\
\widetilde{\alpha}_{\tau e}  & \widetilde{\alpha}_{\tau \mu} & \widetilde{\alpha}_{\tau \tau} \\
\end{array}
\right], ~~ 
\label{alpha-SOL-def}
\end{eqnarray}
where we have introduced the simplified notation $\alpha_{\beta \gamma}^{\text{\tiny SOL}} \equiv \widetilde{\alpha}_{\beta \gamma}$ for our later convenience. 

Therefore, if we talk about the chiral-type correlation between the CP phases in the atmospheric resonance region, it takes the three different forms depending upon the $U$ matrix conventions: $[e^{- i \delta } \alpha_{\mu e}, ~e^{ - i \delta} \alpha_{\tau e}, ~\alpha_{\tau \mu}]$ in the PDG convention, $[e^{- i \delta } \alpha_{\mu e}^{\text{\tiny ATM}}, ~\alpha_{\tau e}^{\text{\tiny ATM}}, ~e^{i \delta} \alpha_{\tau \mu}^{\text{\tiny ATM}}]$ in the ATM convention, and $[\widetilde{\alpha}_{\mu e}, \widetilde{\alpha}_{\tau e}, \widetilde{\alpha}_{\tau \mu}]$ in the SOL convention. 
That is, the phase correlation disappears in the SOL convention. It happens by accident in which the convention dependent change in the $\alpha$ parameters just absorbs the physical phase correlation existed in the PDG and ATM conventions~\cite{Martinez-Soler:2018lcy}. 
Therefore, to our understanding, the phase correlation generically exists and represents the unique physical characteristics of the $\nu$SM and UV phases. No phase convention of the $U$ matrix exists which eliminates the phase correlations at everywhere in the whole kinematical phase space, as discussed in depth in ref.~\cite{Martinez-Soler:2019noy}.

\subsection{We use the SOL convention}
\label{sec:use-SOL}

Vanishing $e^{ \pm i \delta }$ - UV phase correlation in the atmospheric resonance region in the SOL convention suggests that the phase correlation takes the simplest form in wider kinematical phase space in this convention. Therefore, we use the SOL convention for our investigation of the CP phase correlation in the DMP-UV perturbation theory with the notation $\alpha^{\text{\tiny SOL}} \equiv \widetilde{\alpha}$ for simplicity as defined in eq.~\eqref{alpha-SOL-def}. 
One must know that the oscillation probability computed with the use of the PDG, ATM, and the SOL conventions is exactly identical. It is because neutrino-state phase redefinition cannot alter physical observables. To transform the probability $P(\nu_{\beta} \rightarrow \nu_{\alpha})$ using the SOL convention to the one in the PDG, or ATM conventions, however, one must transform the $\widetilde{\alpha}_{\beta \gamma}$ parameters to $\alpha_{\beta \gamma}$ defined in eq.~\eqref{alpha-matrix}, or $\alpha_{\beta \gamma}^{\text{\tiny ATM}}$, respectively, following the translation rules in eqs.~\eqref{alpha-SOL-def} and \eqref{alpha-ATM-def}. 

The SOL convention is used in ref.~\cite{Martinez-Soler:2019noy} to demonstrate physical nature of the CP phase correlations. It is also used in much more crucial way in ref.~\cite{Minakata:2021dqh} for the transparent formulation of ``Symmetry Finder'', a powerful tool for symmetry hunting in neutrino oscillations in matter. 

\section{Computation of the probability $P(\nu_{\mu} \rightarrow \nu_{e})$} 
\label{sec:probability-mue} 

To focus in on the physics discussions on the features of $\nu$SM - UV phase correlations in the main text, we send our compact formulation of the DMP-UV perturbation theory to Appendix~\ref{sec:DMP-UV-formulation}. In this paper, for simplicity and clarity, we compute the probability to first order in DMP-UV expansion, and will work with the uniform matter density approximation.   

\subsection{Preface toward presenting the first-order UV corrections and notations}
\label{sec:preface}

The DMP-UV perturbation theory has two kind of the expansion parameters, $\epsilon$ and the UV $\widetilde{\alpha}$ parameters. $\epsilon$ in the $\nu$SM part is defined as 
\begin{eqnarray} 
&&
\epsilon \equiv \frac{ \Delta m^2_{21} }{ \Delta m^2_{ \text{ren} } }, 
\hspace{10mm}
\Delta m^2_{ \text{ren} } \equiv \Delta m^2_{31} - s^2_{12} \Delta m^2_{21},
\label{epsilon-Dm2-ren-def}
\end{eqnarray}
where $\Delta m^2_{ \text{ren} }$ is the ``renormalized'' atmospheric $\Delta m^2$ used in ref.~\cite{Minakata:2015gra}.\footnote{
The same quantity is known as the effective $\Delta m^2_{ \text{ee} }$ in the $\nu_{e} \rightarrow \nu_{e}$ channel in vacuum~\cite{Nunokawa:2005nx}. While we prefer usage of $\Delta m^2_{ \text{ren} }$ in the context of the present paper, the question of which symbol should be appropriate here is under debate~\cite{Minakata:2015gra}. }
In DMP the vacuum mixing angles are elevated to the matter-dressed one, $\theta_{12} \rightarrow \psi$ and $\theta_{13} \rightarrow \phi$, but $\theta_{23}$ and $\delta$ are as they are in vacuum~\cite{Denton:2016wmg}. 
In the expressions of the oscillation probability, we use the various simplified notations as follow: 
\begin{eqnarray} 
&&
h_{i} \equiv \frac{ \lambda_{i} }{ 2E }, 
\hspace{10mm}
\Delta_{ \text{ren} } \equiv \frac{ \Delta m^2_{ \text{ren} } }{ 2E }, 
\hspace{10mm}
\Delta_{a} \equiv \frac{ a }{ 2E }, 
\hspace{10mm}
\Delta_{b} \equiv \frac{ b }{ 2E }, 
\label{ratios-def}
\end{eqnarray}
$h_{i}$ ($i=1,2,3$) denote the eigenvalues of the unperturbed Hamiltonian, the diagonal entries of $\bar{H}_\text{$\nu$SM}$ in eqs.~\eqref{barH-SM} or \eqref{barH-summary}. $a$ and $b$ are the Wolfenstein matter potentials defined in eq.~\eqref{matt-potential}. We also use $J_{mr} \equiv c_{23} s_{23} c^2_{\phi} s_{\phi} c_{\psi} s_{\psi}$ as the Jarlskog factor~\cite{Jarlskog:1985ht} in matter. 

\subsection{Structure of the $S$ matrix and the probability in DMP-UV perturbation theory}
\label{sec:structure}

We decompose the flavor-basis $S$ matrix into the $\nu$SM part, $S_{\nu\text{SM} } = S_{\nu\text{SM} }^{(0)} + S_{\nu\text{SM} }^{(1)}$, and the UV $\alpha$ parameter dependent part. The latter terms are all first order by definition and consist of the two terms $S_{\text{EV} }^{(1)}$ and $S_{\text{UV} }^{(1)}$. The subscripts ``EV'' and ``UV'' imply the $\alpha$ parameter dependent but unitary evolution part, and the genuine non-unitary contribution, respectively. The latter comes from non-unitary projections at the production and detection points, and the non-unitarity arises solely from it~\cite{Martinez-Soler:2018lcy}. One can show that, 
\begin{eqnarray} 
&&
S_{\text{UV} }^{(1)} 
= 
- \widetilde{\alpha} S_{\nu\text{SM} }^{(0)}
- S_{\nu\text{SM} }^{(0)} \widetilde{\alpha}^{\dagger}.
\label{S-UV}
\end{eqnarray}
See eq.~\eqref{S-flavor-bar}, 
Therefore, the flavor basis $S$ matrix can be written to first order as 
\begin{eqnarray} 
S_{ \text{flavor} } &=& 
S_{\nu\text{SM} }^{(0)} + S_{\nu\text{SM} }^{(1)} + S_{\text{EV} }^{(1)}
- \widetilde{\alpha} S_{\nu\text{SM} }^{(0)}
- S_{\nu\text{SM} }^{(0)} \widetilde{\alpha}^{\dagger}. 
\label{S-flavor-all}
\end{eqnarray}

Now, we are ready to calculate the expressions of the oscillation probability to first order in
the expansion parameters. Following ref.~\cite{Martinez-Soler:2018lcy}, we categorize $P(\nu_{\beta} \rightarrow \nu_{\alpha})$ into the three types of terms:
\begin{eqnarray} 
P(\nu_{\beta} \rightarrow \nu_{\alpha}) &=& 
P(\nu_{\beta} \rightarrow \nu_{\alpha})_{\nu\text{SM} }^{(0+1)} 
+ P(\nu_{\beta} \rightarrow \nu_{\alpha})_{ \text{ EV } }^{(1)} 
+ P(\nu_{\beta} \rightarrow \nu_{\alpha})_{ \text{ UV } }^{(1)}, 
\label{S-flavor-all}
\end{eqnarray}
where 
\begin{eqnarray} 
P(\nu_{\beta} \rightarrow \nu_{\alpha})_{\nu\text{SM} }^{(0+1)} 
&=& 
\biggl| \left( S_{\nu\text{SM} }^{(0)} \right)_{\alpha \beta} \biggr|^2 
+ 2 \mbox{Re} \left[
\left( S_{\nu\text{SM} }^{(0)} \right)_{\alpha \beta}^* 
\left( S_{\nu\text{SM} }^{(1)} \right)_{\alpha \beta} 
\right], 
\nonumber \\
P(\nu_{\beta} \rightarrow \nu_{\alpha})_{ \text{ EV } }^{(1)}
&=& 
2 \mbox{Re} \left[
\left( S_{\nu\text{SM} }^{(0)} \right)_{\alpha \beta}^* 
\left( S^{(1)}_{ \text{ EV } } \right)_{\alpha \beta} 
\right], 
\nonumber \\
P(\nu_{\beta} \rightarrow \nu_{\alpha})_{ \text{ UV } }^{(1)}
&=& 
- 2 \mbox{Re} \left[
\left( S_{\nu\text{SM} }^{(0)} \right)_{\alpha \beta}^* 
\left( \widetilde{\alpha} S_{\nu\text{SM} }^{(0)} + S_{\nu\text{SM} }^{(0)} \widetilde{\alpha}^{\dagger} \right)_{\alpha \beta} 
\right]. 
\label{P-three-types}
\end{eqnarray}
At the end of Appendix~\ref{sec:DMP-UV-formulation}, we will obtain the zeroth and first order expressions of the flavor-basis $S$ matrix. The $\nu$SM part of the oscillation probability $P(\nu_{\beta} \rightarrow \nu_{\alpha})_{\nu\text{SM} }^{(0+1)}$ to first order in $\epsilon$ is fully calculated in ref.~\cite{Denton:2016wmg}, and therefore we do not repeat. The accuracy of the first order formulas $P(\nu_{\beta} \rightarrow \nu_{\alpha})_{\nu\text{SM} }^{(0+1)}$ are verified in ref.~\cite{Parke:2019vbs}. For the explicit expressions of the $\nu$SM part of the probabilities, we refer arXiv v3 of ref.~\cite{Minakata:2020oxb} for the formulas in all the relevant channels. Since our interest in this paper is on the $\nu$SM - UV phase correlations, we focus on $P(\nu_{\beta} \rightarrow \nu_{\alpha})_{ \text{ EV } }^{(1)}$ and $P(\nu_{\beta} \rightarrow \nu_{\alpha})_{ \text{ UV } }^{(1)}$, hereafter.\footnote{
One may ask why we do not include the second order UV effect. We assume, as the objective of our investigation, ``a small but robust'' evidence for departure from the $\nu$SM as mentioned in Introduction. In diagnosing non-unitarity via $S_{\nu\text{SM}}^{*} S_{ \text{UV} }$, the leading-order effect is from the first order terms of $S_{ \text{UV} }$. If we assume $\sim$1\% level non-unitarity, the second order corrections must be negligible. If the UV effect turns out to be $\sim$10\% level it would be a suggestion for revising the zeroth-order paradigm. }

In this paper we only compute the probability in the $\nu_{\mu} \rightarrow \nu_{e}$ channel for relative simplicity. In fact, all the $\widetilde{\alpha}$ parameters show up in $P(\nu_{\mu} \rightarrow \nu_{e})^{(1)}$. From our experience in refs.~\cite{Martinez-Soler:2018lcy,Martinez-Soler:2019noy}, we are confident that this channel is sufficient to extract the qualitative features of the $\nu$SM - UV phase correlations, for which more comments will follow in section~\ref{sec:completed-picture}. To back this statement up, we provide the $\widetilde{\alpha}$ parameter dependent part of the flavor basis $S$ matrix in the $\nu_{\mu} \rightarrow \nu_{\tau}$ channel in Appendix~\ref{sec:S-taumu}, which indeed supports the above statement. Of course, if any demand exists, it is straightforward to compute the probabilities $P(\nu_{\mu} \rightarrow \nu_{\mu})$ or $P(\nu_{\mu} \rightarrow \nu_{\tau})$ by using the formulas we provide in this paper.

The probability expressions contain the $K$ matrix elements $K_{ij}$ (for the definition see eq.~\eqref{K-def}), which can be written by using the $\widetilde{\alpha}$ parameters as 
\begin{eqnarray} 
K_{11} &=& 
2 c^2_{\phi} \widetilde{\alpha}_{ee} \left( 1 - \frac{ \Delta_{a} }{ \Delta_{b} } \right) 
+ 2 s^2_{\phi} 
\left[
s_{23}^2 \widetilde{\alpha}_{\mu \mu} + c_{23}^2 \widetilde{\alpha}_{\tau \tau} + c_{23} s_{23} \mbox{Re} \left( \widetilde{\alpha}_{\tau \mu} \right) 
\right] 
\nonumber \\
&-& 
2 c_{\phi} s_{\phi} 
\mbox{Re} \left( s_{23} \widetilde{\alpha}_{\mu e} + c_{23} \widetilde{\alpha}_{\tau e} \right), 
\nonumber \\
K_{12} &=& 
c_{\phi} \left( c_{23} \widetilde{\alpha}_{\mu e}^* - s_{23} \widetilde{\alpha}_{\tau e}^* \right) 
- s_{\phi} 
\left[ 2 c_{23} s_{23} ( \widetilde{\alpha}_{\mu \mu} - \widetilde{\alpha}_{\tau \tau} ) + c_{23}^2 \widetilde{\alpha}_{\tau \mu} - s_{23}^2 \widetilde{\alpha}_{\tau \mu}^* \right] 
= \left( K_{21} \right)^*, 
\nonumber \\
K_{13} &=& 
2 c_{\phi} s_{\phi} 
\left[
\widetilde{\alpha}_{ee} \left( 1 - \frac{ \Delta_{a} }{ \Delta_{b} } \right) 
- \left( s_{23}^2 \widetilde{\alpha}_{\mu \mu} + c_{23}^2 \widetilde{\alpha}_{\tau \tau} \right) 
\right]
\nonumber \\
&+& 
c^2_{\phi} \left( s_{23} \widetilde{\alpha}_{\mu e}^* + c_{23} \widetilde{\alpha}_{\tau e}^* \right) 
- s^2_{\phi} \left( s_{23} \widetilde{\alpha}_{\mu e} + c_{23} \widetilde{\alpha}_{\tau e} \right) 
- 2 c_{23} s_{23} c_{\phi} s_{\phi} 
\mbox{Re} \left( \widetilde{\alpha}_{\tau \mu} \right) 
= \left( K_{31} \right)^*, 
\nonumber \\
K_{22} &=& 
2 \left[ 
c_{23}^2 \widetilde{\alpha}_{\mu \mu} + s_{23}^2 \widetilde{\alpha}_{\tau \tau} - c_{23} s_{23} \mbox{Re} \left( \widetilde{\alpha}_{\tau \mu} \right) 
\right], 
\nonumber \\
K_{23} &=& 
s_{\phi} \left( c_{23} \widetilde{\alpha}_{\mu e} - s_{23} \widetilde{\alpha}_{\tau e} \right) 
+ c_{\phi} 
\left[ 2 c_{23} s_{23} ( \widetilde{\alpha}_{\mu \mu} - \widetilde{\alpha}_{\tau \tau} ) + c_{23}^2 \widetilde{\alpha}_{\tau \mu}^* - s_{23}^2 \widetilde{\alpha}_{\tau \mu} \right] 
= \left( K_{32} \right)^*, 
\nonumber \\
K_{33} &=& 
2 s^2_{\phi} \widetilde{\alpha}_{ee} \left( 1 - \frac{ \Delta_{a} }{ \Delta_{b} } \right) 
+ 2 c^2_{\phi} 
\left[
s_{23}^2 \widetilde{\alpha}_{\mu \mu} + c_{23}^2 \widetilde{\alpha}_{\tau \tau} + c_{23} s_{23} \mbox{Re} \left( \widetilde{\alpha}_{\tau \mu} \right) 
\right] 
\nonumber \\
&+&
2 c_{\phi} s_{\phi} \mbox{Re} 
\left( s_{23} \widetilde{\alpha}_{\mu e} + c_{23} \widetilde{\alpha}_{\tau e} \right). 
\label{Kij-elements}
\end{eqnarray}

\subsection{The probability $P(\nu_{\mu} \rightarrow \nu_{e})$: Unitary evolution part} 
\label{sec:Pmue-EV}

For convenience, we decompose the unitary evolution part of the probability into the three pieces 
\begin{eqnarray} 
&& 
P(\nu_{\mu} \rightarrow \nu_{e})^{(1)} _{ \text{EV} } 
= 
P(\nu_{\mu} \rightarrow \nu_{e})^{(1)} _{ \text{EV} } \vert_{( \Delta_{b} x )}
+ P(\nu_{\mu} \rightarrow \nu_{e})^{(1)} _{ \text{EV} } \vert_{K_{11, 12, 22} } 
+ P(\nu_{\mu} \rightarrow \nu_{e})^{(1)} _{ \text{EV} } \vert_{K_{13, 23} } 
\nonumber \\
\label{P-mue-EV}
\end{eqnarray}
depending upon the form of the kinematical factors $( \Delta_{b} x )$ or $\frac{ \Delta_{b} }{ h_{2} - h_{1} }$ etc. and on the $K$ matrix elements involved. Here, we note that $x$ denotes the detection point of neutrinos measured from production point $x=0$ so that $x$ implies the baseline. 

The first term in eq.~\eqref{P-mue-EV} is given by 
\begin{eqnarray} 
&& 
P(\nu_{\mu} \rightarrow \nu_{e})^{(1)} _{ \text{EV} } \vert_{( \Delta_{b} x )} 
\nonumber \\
&=& 
2 ( \Delta_{b} x ) 
\biggl[ 
\left\{ \cos 2\psi \left( K_{22} - K_{11} \right) 
+ \sin 2\psi \left( e^{ i \delta} K_{21} + e^{ - i \delta} K_{12} \right) \right\} 
\nonumber \\
&\times&
\biggl\{
\left[ c^2_{\phi} c^2_\psi s^2_\psi 
\left( c^2_{23} - s^2_{23} s^2_{\phi} \right) + J_{mr} \cos \delta \cos 2\psi \right] 
\sin ( h_{2} - h_{1} ) x 
+ 2 J_{mr} \sin \delta 
\sin^2 \frac{ ( h_{2} - h_{1} ) x }{2} 
\biggr\}
\nonumber \\
&-& 
\left\{ \left( s^2_\psi K_{11} + c^2_\psi K_{22} - K_{33} \right) 
+ c_\psi s_\psi \left( e^{ i \delta} K_{21} + e^{ - i \delta} K_{12} \right) \right\} 
\nonumber \\
&\times&
\biggl\{ 
\left[ s^2_{23} c^2_{\phi} s^2_{\phi} s^2_\psi - J_{mr} \cos \delta \right] 
\sin ( h_{3} - h_{2} ) x 
+ 2 J_{mr} \sin \delta 
\sin^2 \frac{ ( h_{3} - h_{2} ) x }{2} 
\biggr\} 
\nonumber \\
&-& 
\left\{ \left( c^2_\psi K_{11} + s^2_\psi K_{22} - K_{33} \right) 
- c_\psi s_\psi \left( e^{ i \delta} K_{21} + e^{ - i \delta} K_{12} \right) \right\} 
\nonumber \\
&\times&
\biggl\{ 
\left[ s^2_{23} c^2_{\phi} s^2_{\phi} c^2_\psi 
+ J_{mr} \cos \delta \right] \sin ( h_{3} - h_{1} ) x 
- 2 J_{mr} \sin \delta 
\sin^2 \frac{ ( h_{3} - h_{1} ) x }{2} 
\biggr\} 
\biggr]. 
\label{P-mue-EV-1st} 
\end{eqnarray}
We should note here that $e^{ i \delta} K_{21} + e^{ - i \delta} K_{12}$ is a real number. The second term in eq.~\eqref{P-mue-EV} reads 
\begin{eqnarray} 
&& 
P(\nu_{\mu} \rightarrow \nu_{e})^{(1)} _{ \text{EV} } \vert_{K_{11, 12, 22} }
\nonumber \\
&=&
2 \biggl[ 
\sin 2\theta_{23} c^2_{\phi} s_{\phi} 
\left\{ \cos \delta \mbox{Re} \left( e^{ - i \delta} K_{12} \right) 
- \sin \delta \mbox{Im} \left( e^{ - i \delta} K_{12} \right) \right\} 
\nonumber \\
&-& 
s^2_{23} c^2_{\phi} s^2_{\phi} \sin 2\psi 
\left\{ \sin 2\psi \left( K_{11} - K_{22} \right) 
+ \cos 2\psi \left( e^{ i \delta} K_{21} + e^{ - i \delta} K_{12} \right) \right\} 
\nonumber \\
&+& 
2 J_{mr} \cos \delta 
\left\{\cos 2\psi \left( K_{11} - K_{22} \right) 
- \sin 2\psi \left( e^{ i \delta} K_{21} + e^{ - i \delta} K_{12} \right) \right\} 
\biggr] 
\nonumber \\
&\times& 
\frac{ \Delta_{b} }{ h_{2} - h_{1} } 
\left\{ - \sin^2 \frac{ ( h_{3} - h_{2} ) x }{2} + \sin^2 \frac{ ( h_{3} - h_{1} ) x }{2} \right\} 
\nonumber \\
&+& 
2 \biggl[
2 c_{23} \left\{ c^2_{\phi} 
\left( c_{23} \sin 2\psi + s_{23} s_{\phi} \cos 2\psi \cos \delta \right) 
\mbox{Re} \left( e^{ - i \delta} K_{12} \right) 
- s_{23} c^2_{\phi} s_{\phi} \cos 2\psi \sin \delta \mbox{Im} \left( e^{ - i \delta} K_{12} \right) \right\} 
\nonumber \\
&+& 
c^2_{23} c^2_{\phi} \sin^2 2\psi
\left\{
\cos 2\psi \left( K_{11} - K_{22} \right) 
- \sin 2\psi \left( e^{ i \delta} K_{21} + e^{ - i \delta} K_{12} \right) 
\right\} 
\nonumber \\
&-& 
s^2_{23} c^2_{\phi} s^2_{\phi} \cos 2\psi \sin 2\psi 
\left\{ \sin 2\psi \left( K_{11} - K_{22} \right) 
+ \cos 2\psi \left( e^{ i \delta} K_{21} + e^{ - i \delta} K_{12} \right) \right\} 
\nonumber \\
&+& 
2 J_{mr} \cos \delta 
\left\{ \cos 4\psi \left( K_{11} - K_{22} \right) 
- \sin 4\psi \left( e^{ i \delta} K_{21} + e^{ - i \delta} K_{12} \right) 
\right\} 
\biggr] 
\frac{ \Delta_{b} }{ h_{2} - h_{1} } \sin^2 \frac{ ( h_{2} - h_{1} ) x }{2} 
\nonumber \\
&-& 
4 \biggl[
\sin 2\theta_{23} c^2_{\phi} s_{\phi} 
\left\{ \sin \delta \mbox{Re} \left( e^{ - i \delta} K_{12} \right) 
+ \cos \delta \mbox{Im} \left( e^{ - i \delta} K_{12} \right) \right\} 
\nonumber \\
&+& 
2 J_{mr} \sin \delta 
\left\{\cos 2\psi \left( K_{11} - K_{22} \right) 
- \sin 2\psi \left( e^{ i \delta} K_{21} + e^{ - i \delta} K_{12} \right) \right\} 
\biggr] 
\nonumber \\
&\times& 
\frac{ \Delta_{b} }{ h_{2} - h_{1} } 
\sin \frac{ ( h_{3} - h_{1} ) x }{2} \sin \frac{ ( h_{2} - h_{1} ) x }{2} 
\sin \frac{ ( h_{3} - h_{2} ) x }{2}. 
\label{P-mue-EV-2nd}
\end{eqnarray} 
In the last line we have used the trigonometric identity 
\begin{eqnarray} 
&& 
\hspace{-6mm}
\left[ \sin ( h_{3} - h_{2} ) x - \sin ( h_{3} - h_{1} ) x 
+ \sin ( h_{2} - h_{1} ) x \right] 
= 4 \sin \frac{ ( h_{3} - h_{1} ) x }{2} \sin \frac{ ( h_{2} - h_{1} ) x }{2} 
\sin \frac{ ( h_{3} - h_{2} ) x }{2}.
\nonumber 
\end{eqnarray}

Here, we need to make a remark about how the $\delta$ dependences is organized in the above formula and its relevance for our later discussion. We are talking about the first line in eq.~\eqref{P-mue-EV-2nd}, the term inside $\left\{ \cdot \cdot \cdot \right\}$, which originates from $K_{12} = e^{ i \delta} \left( e^{ - i \delta} K_{12} \right)$ through which an artificial $\delta$ dependence may have been created. Later we will encounter the same situation for $K_{32} = e^{ i \delta} \left( e^{ - i \delta} K_{32} \right)$. The $K_{12}$ term exists in a term in $( S^{(0)}_{e \mu} )^* S^{(1)}_{e \mu}$, 
\begin{eqnarray} 
&& 
\biggl\{
c_{23} c_{\phi} c_\psi s_\psi e^{ - i \delta} 
\left( e^{ i h_{2} x }  - e^{ i h_{1} x } \right) 
+ s_{23} c_{\phi} s_{\phi} 
\left[ e^{ i h_{3} x } - \left( c^2_\psi e^{ i h_{1} x } + s^2_\psi e^{ i h_{2} x } \right) 
\right] 
\biggr\} 
\nonumber \\&\times& 
c_{23} c_{\phi} K_{12} 
\frac{ \Delta_{b} }{ h_{2} - h_{1} } \left\{ e^{ - i h_{2} x } - e^{ - i h_{1} x } \right\} 
\nonumber \\
&=& 
4 c^2_{23} c^2_{\phi} c_\psi s_\psi \left( e^{ - i \delta} K_{12} \right) 
\frac{ \Delta_{b} }{ h_{2} - h_{1} } 
\sin^2 \frac{ ( h_{2} - h_{1} ) x }{2} 
\nonumber \\
&& 
\hspace{-12mm}
+ c_{23} s_{23} c^2_{\phi} s_{\phi} e^{ i \delta} \left( e^{ - i \delta} K_{12} \right)
\frac{ \Delta_{b} }{ h_{2} - h_{1} } 
\biggl[
2 \left\{ - \sin^2 \frac{ ( h_{3} - h_{2} ) x }{2} 
+ \sin^2 \frac{ ( h_{3} - h_{1} ) x }{2} 
+ \cos 2\psi \sin^2 \frac{ ( h_{2} - h_{1} ) x }{2} \right\} 
\nonumber \\
&& 
\hspace{46mm}
+ i \left[ \sin ( h_{3} - h_{2} ) x - \sin ( h_{3} - h_{1} ) x + \sin ( h_{2} - h_{1} ) x \right] 
\biggr].
\label{Int-3}
\end{eqnarray}
See the last two lines in eq.~\eqref{Int-3}. It leads to the several terms in eq.~\eqref{P-mue-EV-2nd} with the explicit $\cos \delta$ and $\sin \delta$ dependencies, in the first line, fifth line, and the third line from the bottom. 
Though it may look artificial, we believe that this organization is natural. If one go through the results of $P(\nu_{\mu} \rightarrow \nu_{e})^{(1)} _{ \text{EV} }$ in eq.~\eqref{P-mue-EV-2nd}, $\left( e^{ i \delta} K_{21} + e^{ - i \delta} K_{12} \right) = 2 \mbox{Re} \left( e^{ - i \delta} K_{12} \right)$ structure is everywhere, and therefore, the $\delta$-UV correlation of the type $\left( e^{ - i \delta} K_{12} \right)$ is unmistakable. Notice that since $\delta$ lives in the $\nu$SM unperturbed pert of the DMP perturbation theory, as can be seen in eq.~\eqref{flavorS0}, $\delta$ can sneak into every part of the first order corrections. 
However, we must note that in the limit to the atmospheric resonance region the term under discussion survives as one of the $\delta$-independent terms, as will be seen in section~\ref{sec:limit-to-ATM}. 

For bookkeeping purpose we divide the last term (third term) in eq.~\eqref{P-mue-EV} into the two pieces, one proportional to $\frac{ \Delta_{b} }{ h_{3} - h_{1} }$, and the other to $\frac{ \Delta_{b} }{ h_{3} - h_{2} }$. They read 
\begin{eqnarray} 
&& 
P(\nu_{\mu} \rightarrow \nu_{e})^{(1)} _{ \text{EV} } \vert_{K_{13, 23} } 
\vert^{\frac{ \Delta_{b} }{ h_{3} - h_{1} } }
\nonumber \\
&=& 
4 c_{23} c_{\phi} c_\psi s_\psi 
\biggl[ 
- c_{23} s_{\phi} s_\psi 
\mbox{Re} \left( c_\psi K_{31} - s_\psi e^{ - i \delta} K_{32} \right) 
\nonumber \\
&+& 
s_{23} \cos \delta 
\left\{ c^2_{\psi} \mbox{Re} \left( c^2_{\phi} K_{13} - s^2_{\phi} K_{31} \right) 
- c_\psi s_\psi \mbox{Re} \left( c^2_{\phi} e^{ i \delta} K_{23} - s^2_{\phi} e^{ - i \delta} K_{32} \right) \right\} 
\nonumber \\
&+& 
s_{23} \sin \delta 
\left\{ c^2_{\psi} \mbox{Im} \left( c^2_{\phi} K_{13} - s^2_{\phi} K_{31} \right) 
- c_\psi s_\psi \mbox{Im} \left( c^2_{\phi} e^{ i \delta} K_{23} - s^2_{\phi} e^{ - i \delta} K_{32} \right) \right\} 
\biggr] 
\nonumber \\
&\times& 
\frac{ \Delta_{b} }{ h_{3} - h_{1} } 
\left\{ - \sin^2 \frac{ ( h_{3} - h_{2} ) x }{2} 
+ \sin^2 \frac{ ( h_{3} - h_{1} ) x }{2} 
+ \sin^2 \frac{ ( h_{2} - h_{1} ) x }{2} \right\} 
\nonumber \\
&+& 
4 s_{23} c_{\phi} s_{\phi} 
\biggl[
- c_{23} s_{\phi} s_\psi 
\left\{ \cos \delta \mbox{Re} \left( c_\psi K_{31} - s_\psi e^{ - i \delta} K_{32} \right) 
- \sin \delta \mbox{Im} \left( c_\psi K_{31} - s_\psi e^{ - i \delta} K_{32} \right) \right\} 
\nonumber \\
&+& 
s_{23} 
\left\{ c^2_{\psi} \mbox{Re} \left( c^2_{\phi} K_{13} - s^2_{\phi} K_{31} \right) 
- c_\psi s_\psi \mbox{Re} \left( c^2_{\phi} e^{ i \delta} K_{23} - s^2_{\phi} e^{ - i \delta} K_{32} \right) \right\} 
\biggr]
\nonumber \\
&\times& 
\frac{ \Delta_{b} }{ h_{3} - h_{1} } 
\left\{ ( 1 + c^2_\psi ) \sin^2 \frac{ ( h_{3} - h_{1} ) x }{2} 
+ s^2_\psi \sin^2 \frac{ ( h_{3} - h_{2} ) x }{2} 
- s^2_\psi \sin^2 \frac{ ( h_{2} - h_{1} ) x }{2} \right\} 
\nonumber \\
&+& 
8 \biggl[ 
c_{23} c_{\phi} s_{\phi} s^2_\psi 
\biggl\{ 
s_{23} s_{\phi} s_\psi \sin \delta 
\mbox{Re} \left( c_\psi K_{31} - s_\psi e^{ - i \delta} K_{32} \right) 
\nonumber \\
&&
\hspace{24mm}
+ \left( - c_{23} c_\psi + s_{23} s_{\phi} s_\psi \cos \delta \right) 
\mbox{Im} \left( c_\psi K_{31} - s_\psi e^{ - i \delta} K_{32} \right) 
\biggr\} 
\nonumber \\
&-& 
c_{23} s_{23} c_{\phi} c_\psi s_\psi \sin \delta 
\left\{ c^2_{\psi} \mbox{Re} \left( c^2_{\phi} K_{13} - s^2_{\phi} K_{31} \right) 
- c_\psi s_\psi 
\mbox{Re} \left( c^2_{\phi} e^{ i \delta} K_{23} - s^2_{\phi} e^{ - i \delta} K_{32} \right) \right\} 
\nonumber \\
&+& 
s_{23} c_{\phi} 
\left( - s_{23} s_{\phi} s^2_\psi + c_{23} c_\psi s_\psi \cos \delta \right) 
\left\{ c^2_{\psi} \mbox{Im} \left( c^2_{\phi} K_{13} - s^2_{\phi} K_{31} \right) 
- c_\psi s_\psi \mbox{Im} \left( c^2_{\phi} e^{ i \delta} K_{23} - s^2_{\phi} e^{ - i \delta} K_{32} \right) \right\} 
\biggr] 
\nonumber \\
&\times& 
\frac{ \Delta_{b} }{ h_{3} - h_{1} } 
\sin \frac{ ( h_{3} - h_{1} ) x }{2} \sin \frac{ ( h_{2} - h_{1} ) x }{2} 
\sin \frac{ ( h_{3} - h_{2} ) x }{2}, 
\label{P-mue-EV-3rd-1}
\end{eqnarray}
and 
\begin{eqnarray} 
&& 
P(\nu_{\mu} \rightarrow \nu_{e})^{(1)} _{ \text{EV} } \vert_{K_{13, 23} } 
\vert^{\frac{ \Delta_{b} }{ h_{3} - h_{2} } }
\nonumber \\
&=& 
4 c_{23} c_{\phi} c_\psi s_\psi 
\biggl[
c_{23} s_{\phi} c_\psi 
\mbox{Re} \left( s_\psi K_{31} + c_\psi e^{ - i \delta} K_{32} \right) 
\nonumber \\
&+& 
s_{23} \cos \delta 
\left\{ s^2_{\psi} \mbox{Re} \left( c^2_{\phi} K_{13} - s^2_{\phi} K_{31} \right) 
+ c_\psi s_\psi \mbox{Re} \left( c^2_{\phi} e^{ i \delta} K_{23} - s^2_{\phi} e^{ - i \delta} K_{32} \right) \right\} 
\nonumber \\
&+& 
s_{23} \sin \delta 
\left\{  s^2_{\psi} \mbox{Im} \left( c^2_{\phi} K_{13} - s^2_{\phi} K_{31} \right) 
+ c_\psi s_\psi \mbox{Im} \left( c^2_{\phi} e^{ i \delta} K_{23} - s^2_{\phi} e^{ - i \delta} K_{32} \right) \right\} 
\biggr]
\nonumber \\
&\times& 
\frac{ \Delta_{b} }{ h_{3} - h_{2} } 
\left\{ - \sin^2 \frac{ ( h_{3} - h_{2} ) x }{2} 
+ \sin^2 \frac{ ( h_{3} - h_{1} ) x }{2} 
- \sin^2 \frac{ ( h_{2} - h_{1} ) x }{2} \right\} 
\nonumber \\
&+& 
4 s_{23} c_{\phi} s_{\phi} 
\biggl[ 
c_{23} s_{\phi} c_\psi 
\left\{ \cos \delta \mbox{Re} \left( s_\psi K_{31} + c_\psi e^{ - i \delta} K_{32} \right) 
- \sin \delta \mbox{Im} \left( s_\psi K_{31} + c_\psi e^{ - i \delta} K_{32} \right) \right\} 
\nonumber \\
&+& 
s_{23} 
\left\{ s^2_{\psi} \mbox{Re} \left( c^2_{\phi} K_{13} - s^2_{\phi} K_{31} \right) 
+ c_\psi s_\psi \mbox{Re} \left( c^2_{\phi} e^{ i \delta} K_{23} - s^2_{\phi} e^{ - i \delta} K_{32} \right) \right\} 
\biggr] 
\nonumber \\
&\times& 
\frac{ \Delta_{b} }{ h_{3} - h_{2} } 
\left\{ ( 1 + s^2_\psi ) \sin^2 \frac{ ( h_{3} - h_{2} ) x }{2} 
+ c^2_\psi \sin^2 \frac{ ( h_{3} - h_{1} ) x }{2} 
- c^2_\psi \sin^2 \frac{ ( h_{2} - h_{1} ) x }{2} \right\} 
\nonumber \\
&+& 
8 \biggl[
c_{23} c_{\phi} s_{\phi} c^2_\psi 
\biggl\{ 
s_{23} s_{\phi} c_\psi \sin \delta 
\mbox{Re} \left( s_\psi K_{31} + c_\psi e^{ - i \delta} K_{32} \right) 
\nonumber \\
&&
\hspace{24mm}
+ \left( c_{23} s_\psi + s_{23} s_{\phi} c_\psi \cos \delta \right) 
\mbox{Im} \left( s_\psi K_{31} + c_\psi e^{ - i \delta} K_{32} \right) 
\biggr\} 
\nonumber \\
&-& 
c_{23} s_{23} c_{\phi} c_\psi s_\psi \sin \delta 
\left\{ s^2_{\psi} \mbox{Re} \left( c^2_{\phi} K_{13} - s^2_{\phi} K_{31} \right) 
+ c_\psi s_\psi 
\mbox{Re} \left( c^2_{\phi} e^{ i \delta} K_{23} - s^2_{\phi} e^{ - i \delta} K_{32} \right) \right\}  
\nonumber \\
&+& 
s_{23} c_{\phi} 
\left( s_{23} s_{\phi} c^2_\psi + c_{23} c_\psi s_\psi \cos \delta \right) 
\left\{ s^2_{\psi} \mbox{Im} \left( c^2_{\phi} K_{13} - s^2_{\phi} K_{31} \right) 
+ c_\psi s_\psi \mbox{Im} \left( c^2_{\phi} e^{ i \delta} K_{23} - s^2_{\phi} e^{ - i \delta} K_{32} \right) \right\} 
\biggr] 
\nonumber \\
&\times& 
\frac{ \Delta_{b} }{ h_{3} - h_{2} } 
\sin \frac{ ( h_{3} - h_{1} ) x }{2} \sin \frac{ ( h_{2} - h_{1} ) x }{2} 
\sin \frac{ ( h_{3} - h_{2} ) x }{2}. 
\label{P-mue-EV-3rd-2}
\end{eqnarray}
In deriving eqs.~\eqref{P-mue-EV-3rd-1} and \eqref{P-mue-EV-3rd-2}, we have encountered the similar issue of using $K_{32}$ or $e^{ i \delta} \left( e^{ - i \delta} K_{32} \right)$, which is exactly parallel to the $K_{12}$ vs. $e^{ i \delta} \left( e^{ - i \delta} K_{12} \right)$ issue mentioned after eq.~\eqref{P-mue-EV-2nd}. We prefer to make explicit the $\left( e^{ - i \delta} K_{32} \right)$ correlation for the same reason as for the $\left( e^{ - i \delta} K_{12} \right)$ case.

\subsection{The probability $P(\nu_{\mu} \rightarrow \nu_{e})$: Non-unitary part } 
\label{sec:Pmue-UV}

The non-unitary part of the probability $P(\nu_{\mu} \rightarrow \nu_{e})$ is defined in eq.~\eqref{P-three-types}, and it takes the form 
\begin{eqnarray} 
P(\nu_{\mu} \rightarrow \nu_{e})_{ \text{ UV } }^{(1)}
&=& 
- 2 \mbox{Re} \left[
\left( S_{e \mu}^{(0)} \right)^* 
\left\{ 
\left( \widetilde{\alpha}_{ee} + \widetilde{\alpha}_{\mu \mu} \right) S_{e \mu}^{(0)} 
+ \widetilde{\alpha}_{\mu e}^* S_{e e}^{(0)}  \right\} 
\right] 
\nonumber \\
&=& 
- 2 ( \widetilde{\alpha}_{ee} + \widetilde{\alpha}_{\mu \mu} ) \vert S_{e \mu}^{(0)} \vert^2 
- 2 \mbox{Re} \left[ \widetilde{\alpha}_{\mu e} ( S_{e e}^{(0)} )^* S_{e \mu}^{(0)} \right]. 
\label{P-mue-UV-def} 
\end{eqnarray}
It can be calculated as 
\begin{eqnarray} 
&&
P(\nu_{\mu} \rightarrow \nu_{e})_{ \text{ UV } }^{(1)} = 
- 2 ( \widetilde{\alpha}_{ee} + \widetilde{\alpha}_{\mu \mu} ) 
\nonumber \\
&\times& 
\biggl[ 
s^2_{23} \sin^2 2\phi 
\left\{ c^2_\psi \sin^2 \frac{ ( h_{3} - h_{1} ) x }{2} 
+ s^2_\psi \sin^2 \frac{ ( h_{3} - h_{2} ) x }{2} \right\} 
+ c^2_{\phi} \sin^2 2\psi 
\left( c^2_{23} - s^2_{23} s^2_{\phi} \right) 
\sin^2 \frac{ ( h_{2} - h_{1} ) x }{2} 
\nonumber \\
&+& 
4 J_{mr} \cos \delta 
\left\{ 
\sin^2 \frac{ ( h_{3} - h_{1} ) x }{2} - \sin^2 \frac{ ( h_{3} - h_{2} ) x }{2} 
+ \cos 2\psi \sin^2 \frac{ ( h_{2} - h_{1} ) x }{2} 
\right\} 
\nonumber \\
&-& 
8 J_{mr} \sin \delta 
\sin \frac{ ( h_{3} - h_{1} ) x }{2} \sin \frac{ ( h_{2} - h_{1} ) x }{2} 
\sin \frac{ ( h_{3} - h_{2} ) x }{2} 
\biggr]
\nonumber \\
&+& 
2 c_{23} c_{\phi} \sin 2\psi 
\mbox{Re} \left( \widetilde{\alpha}_{\mu e} e^{ i \delta} \right) 
\biggl[
s^2_{\phi} 
\left\{ \sin^2 \frac{ ( h_{3} - h_{2} ) x }{2} - \sin^2 \frac{ ( h_{3} - h_{1} ) x }{2} \right\} 
+ c^2_{\phi} \cos 2\psi \sin^2 \frac{ ( h_{2} - h_{1} ) x }{2} 
\biggr]
\nonumber \\
&+& 
c_{23} c_{\phi} \sin 2\psi 
\mbox{Im} \left( \widetilde{\alpha}_{\mu e} e^{ i \delta} \right) 
\biggl\{
s^2_{\phi} 
\left[ \sin ( h_{3} - h_{2} ) x - \sin ( h_{3} - h_{1} ) x \right] - c^2_{\phi} \sin ( h_{2} - h_{1} ) x 
\biggr\} 
\nonumber \\
&+& 
s_{23} \sin 2\phi 
\left\{ \cos \delta \mbox{Re} \left( \widetilde{\alpha}_{\mu e} e^{ i \delta} \right) 
+ \sin \delta \mbox{Im} \left( \widetilde{\alpha}_{\mu e} e^{ i \delta} \right) \right\} 
\nonumber \\
&\times& 
\biggl[
2 \cos 2\phi 
\left\{ c^2_\psi \sin^2 \frac{ ( h_{3} - h_{1} ) x }{2} + s^2_\psi \sin^2 \frac{ ( h_{3} - h_{2} ) x }{2} \right\} 
- c^2_{\phi} \sin^2 2\psi 
\sin^2 \frac{ ( h_{2} - h_{1} ) x }{2} 
\biggr] 
\nonumber \\
&+& 
s_{23} \sin 2\phi 
\left\{ \sin \delta \mbox{Re} \left( \widetilde{\alpha}_{\mu e} e^{ i \delta} \right) 
- \cos \delta \mbox{Im} \left( \widetilde{\alpha}_{\mu e} e^{ i \delta} \right) \right\} 
\left[ c^2_\psi \sin ( h_{3} - h_{1} ) x + s^2_\psi \sin ( h_{3} - h_{2} ) x \right].
\label{P-mue-UV-1st}
\end{eqnarray}

In finishing up the computation of the probability in the $\nu_{\mu} \rightarrow \nu_{e}$ channel, we remark that the qualitative feature of the $\nu$SM - UV phase correlations in $P(\nu_{\mu} \rightarrow \nu_{e})^{(1)} _{ \text{EV} }$ (see eqs.~\eqref{P-mue-EV-1st}, \eqref{P-mue-EV-2nd}, \eqref{P-mue-EV-3rd-1}, and \eqref{P-mue-EV-3rd-2}) is between the $K_{ij}$ blobs of $\alpha$ parameters and $e^{ \pm i \delta}$, which is akin to the one observed at around the solar-scale enhanced oscillation~\cite{Martinez-Soler:2019noy}. Whereas, the one in $P(\nu_{\mu} \rightarrow \nu_{e})^{(1)} _{ \text{UV} }$ in eq.~\eqref{P-mue-UV-1st} is a ``chiral type'', but with the SOL convention $\widetilde{\alpha}$ parameters. Though it may look like the one found in ref.~\cite{Martinez-Soler:2018lcy} but that was under the PDG or ATM conventions of the $U$ matrix.

We also remark, in repetition of the one in section~\ref{sec:use-SOL}, that if one wants to obtain the expressions of $P(\nu_{\mu} \rightarrow \nu_{e})^{(1)} _{ \text{EV} }$ and $P(\nu_{\mu} \rightarrow \nu_{e})^{(1)} _{ \text{UV} }$ in the PDG (ATM) convention, the replacement of the $\widetilde{\alpha}_{\beta \gamma}$ parameters by $\alpha_{\beta \gamma}$ ($\alpha_{\beta \gamma}^{\text{\tiny ATM}}$) through the translation rule in eq.~\eqref{alpha-SOL-def} (and eq.~\eqref{alpha-ATM-def}) suffices. 

\section{Taking limit to the atmospheric resonance region}
\label{sec:limit-to-ATM}

The careful readers must be anxious about apparent contradiction between our expressions of $P(\nu_{\mu} \rightarrow \nu_{e})^{(1)} _{ \text{EV} }$ given in eqs.~\eqref{P-mue-EV-1st}, \eqref{P-mue-EV-2nd}, \eqref{P-mue-EV-3rd-1}, and \eqref{P-mue-EV-3rd-2}, and the results obtained in our previous paper~\cite{Martinez-Soler:2018lcy} briefly summarized in section~\ref{sec:now-and-next}. In ref.~\cite{Martinez-Soler:2018lcy}, we have claimed that no correlation exists between $e^{ \pm i \delta}$ and the complex UV $\widetilde{\alpha}$ parameters in the SOL convention, which is in apparent contradiction to the feature we have seen in section~\ref{sec:probability-mue}. We should be able to resolve this puzzle, but a miracle seems to be needed for it. 

To clear this point up, we examine the limit of the DMP-UV perturbation theory to region of the atmospheric-scale enhanced oscillations. In fact, such a limit is already studied in a previous paper~\cite{Minakata:2020oxb} for the $\nu$SM DMP perturbation theory, and the analysis should apply to our case as the UV effect is treated as perturbation. 
The suitable limit to approach to the atmospheric-resonance perturbation theory is to take the operational limit $\epsilon \ll 1$ keeping $\theta_{13}$ and $\phi$ finite, and 
\begin{eqnarray} 
&&
r_{a} \equiv \frac{a}{ \Delta m^2_{ \text{ren} } } 
\simeq 
\frac{ \Delta_{a} }{ \Delta_{31} } 
\left( 1 + s^2_{12} \frac{ \Delta_{21} }{ \Delta_{31} } \right) 
\sim \mathcal{O} (1).
\label{ra-def}
\end{eqnarray}
For convenience, we call this limit as the ``ATM limit''.\footnote{
The nature of the ATM limit is ``operational'' in the sense that $\epsilon \equiv \Delta m^2_{21} / \Delta m^2_{ \text{ren} }$ as defined in eq.~\eqref{epsilon-Dm2-ren-def} is the parameter fixed by nature, and not to vary. But, apparently such a limit is necessary to turn the whole DMP theory~\cite{Denton:2016wmg} to the atmospheric-resonance perturbation theory~\cite{Minakata:2015gra}, whose region of validity is restricted to the one around the enhanced atmospheric-scale oscillations. For example, the high-energy limit $\rho E / \Delta m^2_{ \text{ren} } \gg 1$ where $\rho$ is the matter density, inevitably sends the angle $\phi$ to the asymptotic region, $\sin 2\phi \ll 1$, and hence it cannot be the correct limit to 
the atmospheric-resonance perturbation theory. 
}
The key parameter which describes the theory under the ATM limit  is the mixing angle $\psi$, the matter-dressed $\theta_{12}$. It behaves under the limit as 
\begin{eqnarray}
&& 
\sin 2 \psi 
\simeq
\frac{ \pm 2 \epsilon \sin 2\theta_{12} c_{( \phi - \theta_{13} )} }
{ \left[ 1 + r_{a} - \sqrt{ 1 + r_{a}^2 - 2 r_{a} \cos 2\theta_{13} } \right] } + \mathcal{O} (\epsilon^2),
\nonumber \\
&& 
\cos 2 \psi 
\simeq \mp 1 + \mathcal{O} (\epsilon^2), 
\label{2psi}
\end{eqnarray}
where the upper sign is for the normal mass ordering (NMO), and the lower the inverted mass ordering (IMO). $c_{( \phi - \theta_{13} )} \equiv \cos ( \phi - \theta_{13} )$.

The UV amplitude $S_{ \text{UV} }$ is already first order in the UV parameter, and therefore $\sin 2 \psi$ terms in it are of order $\epsilon^2$, which means that they can be ignored. Therefore, in taking the ATM limit we can set $\sin 2 \psi = 0$ (which also implies $J_{mr}=0$) and $\cos 2 \psi = \mp 1$. It means that $c_\psi = 0$ and $s_\psi = 1$ for the NMO, and $c_\psi = 1$ and $s_\psi = 0$ for the IMO. We discuss below the ATM limit for the IMO case.\footnote{
The simplest interpretation of the probability formulas in ref.~\cite{Martinez-Soler:2018lcy} is that the atmospheric resonance level crossing is between the 1-3 states, which implies the IMO according to the state labeling in DMP. The formulas in ref.~\cite{Martinez-Soler:2018lcy} are valid for the NMO if we interpret the ``$\lambda_{1}$-$\lambda_{3}$ crossing'' as ``$\lambda_{-}$-$\lambda_{+}$ crossing'', as done in the original reference~\cite{Minakata:2015gra}. In the DMP language it corresponds to the ``$\lambda_{2}$-$\lambda_{3}$ crossing'' in NMO. For more details see ref.~\cite{Minakata:2020oxb}. } 

Let us start from the non-unitary part of the probability $P(\nu_{\mu} \rightarrow \nu_{e})_{ \text{ UV } }^{(1)}$ which has a much simpler expression than $P(\nu_{\mu} \rightarrow \nu_{e})_{ \text{ EV } }^{(1)}$. The $e^{ \pm i \delta}$ - UV $\widetilde{\alpha}$ parameter correlation originates from the second term of eq.~\eqref{P-mue-UV-def}, 
$- 2 \mbox{Re} \left[ \widetilde{\alpha}_{\mu e} ( S_{e e}^{(0)} )^* S_{e \mu}^{(0)} \right]$. In the ATM limit the relevant part tends to 
\begin{eqnarray} 
&& 
\widetilde{\alpha}_{\mu e} ( S_{e e}^{(0)} )^* S_{e \mu}^{(0)} 
\nonumber \\
&=& 
\widetilde{\alpha}_{\mu e} 
\biggl[
c_{23} c_{\phi} \sin 2\psi e^{ i \delta} 
\left[
s^2_{\phi} 
\left\{ - \sin^2 \frac{ ( h_{3} - h_{2} ) x }{2} + \sin^2 \frac{ ( h_{3} - h_{1} ) x }{2} \right\} 
- c^2_{\phi} \cos 2\psi \sin^2 \frac{ ( h_{2} - h_{1} ) x }{2} 
\right]
\nonumber \\
&+& 
i c_{23} c_{\phi} c_\psi s_\psi e^{ i \delta} 
\biggl\{
s^2_{\phi} 
\left[ \sin ( h_{3} - h_{2} ) x - \sin ( h_{3} - h_{1} ) x \right] - c^2_{\phi} \sin ( h_{2} - h_{1} ) x 
\biggr\} 
\nonumber \\
&-& 
s_{23} \sin 2\phi \cos 2\phi 
\left\{ c^2_\psi \sin^2 \frac{ ( h_{3} - h_{1} ) x }{2} + s^2_\psi \sin^2 \frac{ ( h_{3} - h_{2} ) x }{2} \right\} 
+ s_{23} c^3_{\phi} s_{\phi} \sin^2 2\psi 
\sin^2 \frac{ ( h_{2} - h_{1} ) x }{2} 
\nonumber \\
&-& 
i s_{23} c_{\phi} s_{\phi} 
\left[ c^2_\psi \sin ( h_{3} - h_{1} ) x + s^2_\psi \sin ( h_{3} - h_{2} ) x \right] 
\biggr]
\nonumber \\
&\rightarrow&_{ \text{ATM} }
\nonumber \\
&=& 
\widetilde{\alpha}_{\mu e} 
\left\{ 
- s_{23} \sin 2\phi \cos 2\phi \sin^2 \frac{ ( h_{3} - h_{1} ) x }{2} 
- i s_{23} c_{\phi} s_{\phi} \sin ( h_{3} - h_{1} ) x
\right\}. 
\label{P-mue-Int-ATM}
\end{eqnarray}
Therefore, the terms with $e^{ \pm i \delta}$ - UV $\widetilde{\alpha}$ parameter correlation $e^{ i \delta} \widetilde{\alpha}_{\mu e}$ vanish, and only the term with $\widetilde{\alpha}_{\mu e}$ remains without $\delta$. In fact, under the ATM limit, $P(\nu_{\mu} \rightarrow \nu_{e})_{ \text{UV} }^{(1)}$ becomes 
\begin{eqnarray} 
&& 
P(\nu_{\mu} \rightarrow \nu_{e})_{ \text{UV} }^{(1)} 
\nonumber \\
&=& 
2 s_{23} \sin 2\phi 
\biggl\{
\cos 2\phi \mbox{Re} \left( \widetilde{\alpha}_{\mu e} \right) 
- s_{23} \sin 2\phi ( \widetilde{\alpha}_{ee} + \widetilde{\alpha}_{\mu \mu} )  
\biggr\} 
\sin^2 \frac{ ( h_{3} - h_{1} ) x }{2} 
\nonumber \\
&-& 
s_{23} \sin 2\phi 
\mbox{Im} \left( \widetilde{\alpha}_{\mu e} \right) 
\sin ( h_{3} - h_{1} ) x, 
\label{P-mue-UV-ATM}
\end{eqnarray}
which reproduces eq.~(50) in ref.~\cite{Martinez-Soler:2018lcy}, under the identifications (see eqs.~\eqref{alpha-SOL-def} and \eqref{alpha-ATM-def}) 
\begin{eqnarray} 
&& 
\widetilde{\alpha}_{\mu e} 
= e^{ - i \delta} \alpha^{\text{\tiny ATM}}_{\mu e}, 
\hspace{10mm}
\widetilde{\alpha}_{\tau e} 
= \alpha^{\text{\tiny ATM}}_{\tau e}, 
\hspace{10mm}
\widetilde{\alpha}_{\tau \mu} 
= e^{ i \delta} \alpha^{\text{\tiny ATM}}_{\tau \mu}.
\label{identification}
\end{eqnarray}


Now, we turn to the unitary evolution part $P(\nu_{\mu} \rightarrow \nu_{e})_{ \text{EV} }^{(1)}$. In the ATM limit all the terms with $e^{ - i \delta} K_{12}$ or $e^{ i \delta} K_{23}$ type correlations disappear, and $P(\nu_{\mu} \rightarrow \nu_{e})_{ \text{EV} }^{(1)}$ approaches to the form 
\begin{eqnarray} 
&& 
P(\nu_{\mu} \rightarrow \nu_{e})_{ \text{EV} }^{(1)} 
\nonumber \\
&=& 
- 2 s^2_{23} c^2_{\phi} s^2_{\phi} ( \Delta_{b} x ) 
\left( K_{11} - K_{33} \right) \sin ( h_{3} - h_{1} ) x 
\nonumber \\
&+& 
2 \sin 2\theta_{23} c^2_{\phi} s_{\phi} 
\mbox{Re} \left( K_{12} \right) 
\frac{ \Delta_{b} }{ h_{2} - h_{1} } 
\left\{ - \sin^2 \frac{ ( h_{3} - h_{2} ) x }{2} 
+ \sin^2 \frac{ ( h_{3} - h_{1} ) x }{2} 
+ \sin^2 \frac{ ( h_{2} - h_{1} ) x }{2} \right\} 
\nonumber \\
&+& 
2 \sin 2\theta_{23} c_{\phi} s^2_{\phi} 
\mbox{Re} \left( K_{32} \right) 
\frac{ \Delta_{b} }{ h_{3} - h_{2} } 
\left\{ 
\sin^2 \frac{ ( h_{3} - h_{2} ) x }{2} 
+ \sin^2 \frac{ ( h_{3} - h_{1} ) x }{2} 
- \sin^2 \frac{ ( h_{2} - h_{1} ) x }{2} 
\right\} 
\nonumber \\
&+& 
4 s^2_{23} \sin 2\phi \cos 2\phi 
\mbox{Re} \left( K_{13} \right) 
\frac{ \Delta_{b} }{ h_{3} - h_{1} } 
\sin^2 \frac{ ( h_{3} - h_{1} ) x }{2} 
\nonumber \\
&-& 
2 \sin 2\theta_{23} \sin 2\phi c_{\phi} 
\mbox{Im} \left( K_{12} \right) 
\frac{ \Delta_{b} }{ h_{2} - h_{1} } 
\sin \frac{ ( h_{3} - h_{1} ) x }{2} \sin \frac{ ( h_{2} - h_{1} ) x }{2} \sin \frac{ ( h_{3} - h_{2} ) x }{2} 
\nonumber \\
&+& 
2 \sin 2\theta_{23} \sin 2\phi s_{\phi} 
\mbox{Im} \left( K_{32} \right) 
\frac{ \Delta_{b} }{ h_{3} - h_{2} } 
\sin \frac{ ( h_{3} - h_{1} ) x }{2} \sin \frac{ ( h_{2} - h_{1} ) x }{2} \sin \frac{ ( h_{3} - h_{2} ) x }{2}. 
\label{P-mue-EV-ATM}
\end{eqnarray}
It simplified dramatically, from the total 41 lines to only 6 lines as in eq.~\eqref{P-mue-EV-ATM} under the ATM limit. Notice that $K_{ij}$ contain only the $\widetilde{\alpha}$ parameters but no $\delta$, see eq.~\eqref{Kij-elements}. Therefore, $\delta$ - $\widetilde{\alpha}$ correlation disappears in $P(\nu_{\mu} \rightarrow \nu_{e})_{ \text{EV} }^{(1)}$ under the ATM limit. 
The expressions of $P(\nu_{\mu} \rightarrow \nu_{e})_{ \text{EV} }^{(1)}$ with explicit usage of the $\alpha$ parameters is given in eq.~\eqref{P-mue-EV-1st-alpha} in Appendix~\ref{sec:P-mue-EV-1st-atm}, which reproduces precisely $P(\nu_{\mu} \rightarrow \nu_{e})_{ \text{EV} }^{(1)}$ in eq.~(49) in ref.~\cite{Martinez-Soler:2018lcy}. 

Thus, we were able to resolve the apparent puzzle. Our results of $P(\nu_{\mu} \rightarrow \nu_{e})_{ \text{EV} }^{(1)}$ and $P(\nu_{\mu} \rightarrow \nu_{e})_{ \text{UV} }^{(1)}$ given in sections~\ref{sec:Pmue-EV} and \ref{sec:Pmue-UV} are miraculously fully consistent with the one in ref.~\cite{Martinez-Soler:2018lcy}. 

\subsection{Another miracle?}
\label{sec:miracle}

The next question we must answer is how wide is the region in which the chiral type correlation $[e^{- i \delta } \alpha_{\mu e}, ~e^{ - i \delta} \alpha_{\tau e}, ~\alpha_{\tau \mu}]$ (PDG convention) exists. As we have learnt in the foregoing treatment in this section, taking the limit to the region $\sin \psi \simeq \epsilon$ (IMO) and $\sin \psi \simeq 1$ (NMO) in DMP guarantees the chiral-type correlation. In view of Fig.~1 in ref.~\cite{Denton:2016wmg} $\sin 2\psi$ is small, $\simeq 2 \epsilon$ ($\psi$ is close to $\pm \pi/2$ or 0), in the entire high-energy region $\vert Y_e \rho E \vert \gsim 3 (\text{g/cm}^3)$ GeV. Since $\psi$ is a monotonically increasing function of $Y_e \rho E$ there is no other region where $\sin 2\psi$ is small. Or, in other word, the (blobs of the $\alpha$ parameters) - $e^{ \pm i \delta}$ correlation can exist only in the region $\vert Y_e \rho E \vert \lsim 3 (\text{g/cm}^3)$ GeV, where $\psi$ undergoes a sharp change from the negative to positive regions of $Y_e \rho E$, i.e., the near vacuum regime. 

Now, we must ask the final question: Is there any possibility that some other limiting procedures bring us to the features of the $\nu$SM - UV CP phase correlations quite different from what we already know? We believe that such another miracle is very unlikely to occur. In pursuing such possibility, we look for a special region of $\phi$, where it jumps or becomes small. But $\phi$ is also monotonically increasing (decreasing) function of $Y_e \rho E$ for the NMO (IMO). Then, our target is practically the regions of small $\sin 2\phi \simeq 2 \epsilon$. In Fig.~1 in ref.~\cite{Denton:2016wmg}, such region does exist in $\vert Y_e \rho E \vert \gsim 40 (\text{g/cm}^3)$ GeV, but it is inside the small $\sin 2\psi$ region where the chiral-type correlation lives. That is, no new feature of phase correlation is expected. We note that this is the region where the all the $\nu$SM oscillation modes die away due to strong matter effect. Nothing interesting happens there for the $\nu$SM oscillations and hence the region does not appear to fit to our purpose of diagnosing non-unitarity through interference. We note that our treatment is valid in the energy region $\vert \rho E \vert \lsim 100 (\text{g/cm}^3)$ GeV, as discussed in section~\ref{sec:UV-vs-sterile}, which is not so far from the region where $\phi \simeq \epsilon$ starts. 

Therefore, we believe that no qualitatively new feature of the $\nu$SM - UV CP phase correlations is expected beyond the two characteristic patterns that have already seen. They are the chiral-type correlation in the atmospheric resonance region, and (blobs of the UV $\alpha$ parameters) - $\delta$ correlations anywhere else. 

\section{A completed picture of $\nu$SM - UV CP phase correlations}
\label{sec:completed-picture} 

Now, we are able to draw a completed picture of the $e^{ \pm i \delta }$ - UV phase correlations in the whole region of the terrestrial neutrino experiments. There exist the two regions in which the characteristically different patterns of the correlations reside: 
\begin{itemize}

\item 
$e^{ \pm i \delta} \alpha_{\beta \gamma}$ chiral-type correlations in region of the atmospheric-scale enhanced oscillations, which extends to higher energies, $\vert \rho E \vert \gsim 6 \text{g/cm}^3$ GeV. 

\item 
$K_{ij}$ (the blobs of the UV $\alpha$ parameters) - $e^{ \pm i \delta}$ correlations everywhere else. 

\end{itemize}
\noindent

Let us summarize the features of the correlation in the latter region, as the former region is discussed in detail in the previous section~\ref{sec:limit-to-ATM}. The characteristic features of $\nu$SM - UV phase correlations in $P(\nu_{\mu} \rightarrow \nu_{e})^{(1)} _{ \text{EV} }$ can be extracted as the three combinations of the $\nu$SM phase $e^{ \pm i \delta}$ factor and the $K_{ij}$ blobs of the $\widetilde{\alpha}$ parameters, $e^{ - i \delta} K_{12}$, $K_{13}$ (no correlation with $\delta$), and $e^{ i \delta} K_{23}$. We recall that the $K_{ij}$ blobs are free from $\delta$ and the explicit expressions of these correlated variables are shown by using eq.~\eqref{Kij-elements} as follows: 
\begin{eqnarray} 
e^{ - i \delta} K_{12} 
&=& 
e^{ - i \delta} 
\biggl\{
c_{\phi} \left( c_{23} \widetilde{\alpha}_{\mu e}^* - s_{23} \widetilde{\alpha}_{\tau e}^* \right) 
- s_{\phi} 
\left[ 2 c_{23} s_{23} ( \widetilde{\alpha}_{\mu \mu} - \widetilde{\alpha}_{\tau \tau} ) + c_{23}^2 \widetilde{\alpha}_{\tau \mu} - s_{23}^2 \widetilde{\alpha}_{\tau \mu}^* \right] 
\biggr\}, 
\nonumber \\
K_{13} &=& 
2 c_{\phi} s_{\phi} 
\left[
\widetilde{\alpha}_{ee} \left( 1 - \frac{ \Delta_{a} }{ \Delta_{b} } \right) 
- \left( s_{23}^2 \widetilde{\alpha}_{\mu \mu} + c_{23}^2 \widetilde{\alpha}_{\tau \tau} \right) 
\right] 
\nonumber \\
&+& 
c^2_{\phi} \left( s_{23} \widetilde{\alpha}_{\mu e}^* + c_{23} \widetilde{\alpha}_{\tau e}^* \right) 
- s^2_{\phi} \left( s_{23} \widetilde{\alpha}_{\mu e} + c_{23} \widetilde{\alpha}_{\tau e} \right) 
- 2 c_{23} s_{23} c_{\phi} s_{\phi} 
\mbox{Re} \left( \widetilde{\alpha}_{\tau \mu} \right), 
\nonumber \\
e^{ i \delta} K_{23} 
&=& 
e^{ i \delta} 
\biggl\{ 
s_{\phi} \left( c_{23} \widetilde{\alpha}_{\mu e} - s_{23} \widetilde{\alpha}_{\tau e} \right) 
+ c_{\phi} 
\left[ 2 c_{23} s_{23} ( \widetilde{\alpha}_{\mu \mu} - \widetilde{\alpha}_{\tau \tau} ) + c_{23}^2 \widetilde{\alpha}_{\tau \mu}^* - s_{23}^2 \widetilde{\alpha}_{\tau \mu} \right] 
\biggr\}.
\label{Kij-correlation}
\end{eqnarray}
Notice that the (blobs of the $\alpha$ parameters) - $e^{ \pm i \delta}$ correlation in $P(\nu_{\mu} \rightarrow \nu_{e})^{(1)} _{ \text{EV} }$ has been seen in region of the solar-scale enhanced oscillation~\cite{Martinez-Soler:2019noy}, we have shown that its region of validity extends to a much wider region satisfying $\vert \rho E \vert \lsim 6 \text{g/cm}^3$ GeV. 

The $\nu$SM - UV phase correlations in $P(\nu_{\mu} \rightarrow \nu_{e})^{(1)} _{ \text{UV} }$ involves only $\widetilde{\alpha}_{\mu e}$ by definition in eq.~\eqref{P-mue-UV-def}. Then, the question is whether this feature is consistent with the above (blobs of the $\alpha$ parameters) - $e^{ \pm i \delta}$ correlation. The answer is yes in the sense that the same $\widetilde{\alpha}_{\mu e} e^{ i \delta}$ correlation as in eq.~\eqref{P-mue-UV-1st} is buried in the blob type correlation in $P(\nu_{\mu} \rightarrow \nu_{e})^{(1)} _{ \text{EV} }$, as seen in eq.~\eqref{Kij-correlation}. Therefore, everything is consistent as far as the (blobs of the UV $\alpha$ parameters) - $e^{ \pm i \delta}$ correlations are concerned. 

In the alternative region $\vert \rho E \vert \gsim 6 \text{g/cm}^3$ GeV, there is no correlations between $e^{ \pm i \delta}$ and the $\widetilde{\alpha}$ parameters in the SOL convention, or the chiral-type $[e^{- i \delta } \alpha_{\mu e}, ~e^{ - i \delta} \alpha_{\tau e}, ~\alpha_{\tau \mu}]$ correlation lives in the PDG convention. This picture applies both to the $P(\nu_{\mu} \rightarrow \nu_{e})^{(1)} _{ \text{EV} }$ and $P(\nu_{\mu} \rightarrow \nu_{e})^{(1)} _{ \text{UV} }$. 

The fact that the above picture of the $\nu$SM - UV phase correlations comes only from the $\nu_{\mu} \rightarrow \nu_{e}$ channel may trigger an obvious question if the feature of the correlations is the same in the other channels. In fact, there are ample supporting evidences for it in our previous exercises~\cite{Martinez-Soler:2018lcy,Martinez-Soler:2019noy}. In addition to these comments it may be worthwhile to add a remark about the $\nu_{\mu} \rightarrow \nu_{\tau}$ channel. As the expression of $P(\nu_{\mu} \rightarrow \nu_{\tau})$ is lengthy, we just write down the flavor basis $S$ matrix of the $\nu_{\mu} \rightarrow \nu_{\tau}$ channel in Appendix~\ref{sec:S-taumu}. From eq.~\eqref{flavor-S-taumu-1st}, it is clear that the above discussed features of $\nu$SM - UV phase correlations prevail in the $\nu_{\mu} \rightarrow \nu_{\tau}$ channel. 

\subsection{Obtaining the $\widetilde{\alpha}$ parameters in region of the blob correlation}
\label{sec:alpha-from-K}

Outside the atmospheric-resonance region, we measure the $K_{ij}$ parameters.\footnote{
It may be interesting to note that the form of $[ e^{ - i \delta} K_{12}, K_{13}, e^{ i \delta} K_{23} ]$ correlation is somewhat reminiscent of the chiral correlation $[ e^{- i \delta } \alpha_{\mu e}, \alpha_{\tau e}, e^{i \delta} \alpha_{\tau \mu} ]$ despite that the latter is in the ATM convention of the $U$ matrix. It suggests an interesting picture that either the $\alpha$ parameters, or the $K$ parameters are the basic elements of the $\nu$SM - UV phase correlations, the $\alpha$- and $K$-parameters duality, which exists in a way bridging between the different $U$ matrix conventions. 
}
In this case, one can invert eq.~\eqref{Kij-elements} for the $\alpha$ parameters as 
$A = U_{23} F U_{23}^{\dagger} = U_{23} U_{13}(\phi) K U^\dagger_{13}(\phi) U_{23}^{\dagger}$, where the $A$ matrix is defined immediately below eq.~\eqref{tilde-H-matt-1st}. The explicit forms of the inverted expressions are given by 
\begin{eqnarray} 
\widetilde{\alpha}_{ee} 
&=&
\frac{1}{2} 
\left( 1 - \frac{ \Delta_{a} }{ \Delta_{b} } \right)^{-1} 
\biggl\{
c^2_{\phi} K_{11} + s^2_{\phi} K_{33} + c_{\phi} s_{\phi} \left( K_{31} + K_{13} \right) 
\biggr\}, 
\nonumber \\
&&
\hspace{-18mm}
\widetilde{\alpha}_{\mu \mu} 
= 
\frac{1}{2}
\left\{
c^2_{23} K_{22} + s^2_{23} 
\left[ s^2_{\phi} K_{11} + c^2_{\phi} K_{33} - c_{\phi} s_{\phi} \left( K_{31} + K_{13} \right) \right] 
- c_{23} s_{23} 
\left[ s_{\phi} \left( K_{12} + K_{21} \right) - c_{\phi} \left( K_{23}  + K_{32} \right) \right] 
\right\}, 
\nonumber \\
&&
\hspace{-18mm}
\widetilde{\alpha}_{\tau \tau} 
= 
\frac{1}{2}
\biggl\{
s^2_{23} K_{22} + c^2_{23} 
\left[ s^2_{\phi} K_{11} + c^2_{\phi} K_{33} - c_{\phi} s_{\phi} \left( K_{31} + K_{13} \right) \right] 
+ c_{23} s_{23} 
\left[ s_{\phi} \left( K_{12} + K_{21} \right) - c_{\phi} \left( K_{23}  + K_{32} \right) \right] 
\biggr\}, 
\nonumber \\
\widetilde{\alpha}_{\mu e} 
&=&
c_{23} \left( c_{\phi} K_{21} + s_{\phi} K_{23} \right) 
+ s_{23} 
\left[ c^2_{\phi} K_{31} - s^2_{\phi} K_{13} + c_{\phi} s_{\phi} \left( K_{33} - K_{11} \right) \right], 
\nonumber \\
\widetilde{\alpha}_{\tau e} 
&=&
- s_{23} \left( c_{\phi} K_{21} + s_{\phi} K_{23} \right) 
+ c_{23} 
\left[ c^2_{\phi} K_{31} - s^2_{\phi} K_{13} + c_{\phi} s_{\phi} \left( K_{33} - K_{11} \right) \right], 
\nonumber \\
&&
\hspace{-18mm}
\widetilde{\alpha}_{\tau \mu} 
= 
s_{\phi} \left( s^2_{23} K_{21} - c^2_{23} K_{12} \right) 
+ c_{\phi} \left( c^2_{23} K_{32} - s^2_{23} K_{23} \right) 
+ c_{23} s_{23} 
\left[ s^2_{\phi} K_{11} + c^2_{\phi} K_{33} - K_{22} - c_{\phi} s_{\phi} \left( K_{31} + K_{13} \right) \right].
\nonumber \\
\label{alpha-by-K}
\end{eqnarray}
Therefore, by assuming measurement with sufficient precision we can determine all the $\widetilde{\alpha}_{\beta \gamma}$, in principle, in the whole region of the terrestrial experiments. 

\subsection{Some remarks about measurement of the $\alpha$ parameters} 
\label{sec:remarks-alpha} 

Probably, the most salient feature of the probability $P(\nu_{\mu} \rightarrow \nu_{e})^{(1)} _{ \text{EV} }$ computed in section~\ref{sec:probability-mue}, as well as the flavor basis $S$ matrix of the $\nu_{\mu} \rightarrow \nu_{\tau}$ channel in Appendix~\ref{sec:S-taumu}, is that all the $K_{ij}$, and hence the $\widetilde{\alpha}_{\beta \gamma}$ ($\beta, \gamma = e, \mu, \tau$) parameters with all possible flavor indices come-in into the expressions. According to our experiences~\cite{Martinez-Soler:2018lcy,Martinez-Soler:2019noy}, this feature remains hold in the other channels as well, and even true in $P(\nu_{e} \rightarrow \nu_{e})^{(1)}$. It means that one-by-one strategy, measuring one parameter at one channel and the other by another channel, does {\em not} work, which implies that we need an extremely high-precision measurement of the probability to determine, or constrain, all the $\widetilde{\alpha}_{\beta \gamma}$ parameters simultaneously. 

To determine nine degrees of freedom of the $\widetilde{\alpha}$ parameters at the same time, practically, one must combine all the available channels by utilizing the accelerator LBL, atmospheric, reactor, and the solar neutrino observations, hopefully with the both neutrino and antineutrino channels if available. We are aware that these requirements are extremely demanding to make experimentally. 
In foreseeing such measurement, it may be worthwhile to examine the $\widetilde{\alpha}_{\beta \gamma}$ dependences of all the probabilities $P(\nu_{\beta} \rightarrow \nu_{\alpha})$ in all the channels. Such an attempt is carried out partially in ref.~\cite{Martinez-Soler:2019noy}, but it must be extended to all the channels. 

\section{Concluding remarks}
\label{sec:conclusion}

After we gave a summary in section~\ref{sec:completed-picture} of what we have learnt about the $\nu$SM - UV phase correlation in this work, only a few remarks are needed to conclude. 

The most salient feature of the $\delta$ - UV phase correlation we have observed is that it simplifies at high energies, $\vert \rho E \vert \gsim 6 \text{g/cm}^3$ GeV. The UV terms are relatively large in this region because it is proportional to the Wolfenstein matter potentials. But, it is highly nontrivial to see that the effect of enhanced UV effect manifests itself in the behavior of $\delta$ - UV phase correlation in the so striking manner, altering the 
(blobs of the $\alpha$ parameters) - $e^{ \pm i \delta}$ correlation to the much simpler ``chiral type'' correlations $[e^{- i \delta } \alpha_{\mu e}, ~e^{ - i \delta} \alpha_{\tau e}, ~\alpha_{\tau \mu}]$ (PDG convention)~\cite{Martinez-Soler:2018lcy}. 
In the remaining region of $\vert \rho E \vert$, however, we have found that the (blobs of the $\alpha$ parameters) - $e^{ \pm i \delta}$ correlation dominates. The globally valid feature may suggest that the $K_{ij}$ parameters are more natural variables to describe the features of the UV at low energies. This feature as well as the possible $\alpha$- and $K$-parameters duality left us with a question, whether the $\alpha$ or the $K_{ij}$ parameters could appear naturally in some UV models. 

An obviously promising strategy for exploring the characteristic feature of the $\delta$ - UV phase correlation would be to sweep over the energy region $E= 0.1$ to 10 GeV, where the key variable $\psi$ which controls the phase correlations has a dramatic change, as shown in  Fig.~1 in ref.~\cite{Denton:2016wmg}. It would require super-precision measurement throughout the region covered by ESSnuSB~\cite{ESSnuSB:2021azq}, T2K-T2HK~\cite{T2K:2021xwb,Hyper-Kamiokande:2018ofw}, NOvA-DUNE~\cite{NOvA:2021nfi,DUNE:2020ypp}, and T2KK\footnote{
A possible acronym used in ref.~\cite{Kajita:2006bt}, but now for the updated name for the setting, ``Tokai-to-Kamioka observatory-Korea neutrino observatory''. } 
\cite{Hyper-Kamiokande:2016srs}. If wide-energy covering atmospheric neutrino measurement has a promising feature for improving precision, Super-~\cite{Super-Kamiokande:2017yvm} and Hyper-Kamiokande~\cite{Hyper-Kamiokande:2018ofw}, DUNE~\cite{DUNE:2020ypp}, as well as IceCube~\cite{IceCubeCollaboration:2021euf} and KM3NeT~\cite{KM3NeT:2021ozk} would become the strong competitors. 
In the realm of natural $K_{ij}$ variables, experimental search for UV may also be pursuit in low-energy LBL set up with the solar-scale enhancement. Possible physics potential in this region is explored in the earlier studies~\cite{Peres:2003wd,Peres:2009xe,Akhmedov:2008qt} and revisited in the recent ones~\cite{Martinez-Soler:2019nhb,Minakata:2019gyw,Yasuda:2020cff,JUNO:2021tll,ESSnuSB:2021azq}. 

We have started to describe our interests in $\nu$SM - UV phase correlations by saying that ultimately we want to construct a machinery to diagnose non-unitarity. A natural question would then be in which way the knowledge of the $\nu$SM - UV phase correlations can merit the diagnosing capability. From experimentalists' view our work may be regarded as a piece for creating ``theorists' analysis code'' in preparation for the real measurement. By knowing the $\delta$ - UV phase correlations we would have a better view of the ``migration matrix'' which describes variable (and their error) correlations in the $\delta$ row. Since the ``determination of all $\alpha_{\beta \gamma}$ at once'' strategy is required for the $\alpha$ parameters, as discussed in section~\ref{sec:remarks-alpha}, the migration matrix is large and therefore any knowledge of its structure should help. 

Thus, even assuming that our discussions in this paper go along the right direction, our phenomenological study in this paper may be regarded as just a starting step from one particular side. It is likely that one should approach to nature of UV from various sides, for example by examining illuminating models of neutrino mass, e.g., in refs.~\cite{Mohapatra:2006gs,Gavela:2009cd} and the fully consistent models of sterile neutrinos, such as in refs.~\cite{Dasgupta:2021ies,Branco:2020yvs}, which may reveal what the $\alpha$ or the $K$ parameters imply in more physical context. 

\appendix

\section{Formulating the DMP perturbation theory with unitarity violation} 
\label{sec:DMP-UV-formulation} 

We present the formulation of the DMP-UV perturbation theory in the SOL convention to make this paper self-contained. It has some overlaps with refs.~\cite{Denton:2016wmg,Minakata:2015gra,Martinez-Soler:2018lcy}, but the full formulation is worth to present because inclusion of UV effect drastically changes the structure of the perturbation theory. 

\subsection{Tilde basis Hamiltonian}
\label{sec:tilde-H}

In consistent with our statement made in section~\ref{sec:3nu-non-unitarity} we start from the Schr\"odinger equation in the vacuum mass eigenstate basis, eq.~\eqref{evolution-check-basis}. In the ``tilde basis'' which is related to the vacuum mass eigenstate basis $\check{\nu}$ as $\widetilde{\nu} = ( U_{13} U_{12} ) \check{\nu}$, the Hamiltonian eq.~\eqref{evolution-check-basis} defined in the check basis is given by using the SOL convention $\widetilde{\alpha}$ matrix as 
\begin{eqnarray} 
&&
\widetilde{H} 
=
( U_{13} U_{12} ) \check{H} ( U_{13} U_{12} )^{\dagger} 
\nonumber \\ 
&=& 
\frac{ \Delta m^2_{ \text{ren} } }{ 2E }
\left\{
\left[
\begin{array}{ccc}
s^2_{13} & 0 & c_{13} s_{13} \\
0 & 0 & 0 \\
c_{13} s_{13}  & 0 & c^2_{13} 
\end{array}
\right] 
+
\epsilon 
\left[
\begin{array}{ccc}
s^2_{12} & 0 & 0 \\
0 & c^2_{12} & 0 \\
0 & 0 & s^2_{12} 
\end{array}
\right] 
+ 
\epsilon c_{12} s_{12} 
\left[
\begin{array}{ccc}
0 & 
c_{13} e^{ i \delta} & 
0 \\
c_{13} e^{ - i \delta} & 
0 & 
- s_{13} e^{ - i \delta} \\
0 & 
- s_{13} e^{ i \delta} & 
0 \\
\end{array}
\right] 
\right\} 
\nonumber \\
&+& 
U_{23}^{\dagger} 
\left\{
{\bf 1} - 
\left[
\begin{array}{ccc}
\widetilde{\alpha}_{ee} & \widetilde{\alpha}_{\mu e}^* & \widetilde{\alpha}_{\tau e}^*  \\
0 & \widetilde{\alpha}_{\mu \mu} & \widetilde{\alpha}_{\tau \mu}^* \\
0 & 0 & \widetilde{\alpha}_{\tau \tau} \\
\end{array}
\right] 
\right\}
\left[
\begin{array}{ccc}
\Delta_{a} - \Delta_{b} & 0 & 0 \\
0 & - \Delta_{b} & 0 \\
0 & 0 & - \Delta_{b} \\
\end{array}
\right] 
\left\{ 
\bf{1} - 
\left[ 
\begin{array}{ccc}
\widetilde{\alpha}_{ee} & 0 & 0 \\
\widetilde{\alpha}_{\mu e} & \widetilde{\alpha}_{\mu \mu}  & 0 \\
\widetilde{\alpha}_{\tau e}  & \widetilde{\alpha}_{\tau \mu} & \widetilde{\alpha}_{\tau \tau} \\
\end{array}
\right] 
\right\}
U_{23} 
\nonumber \\ 
&\equiv& 
\widetilde{H}_\text{vac} + \widetilde{H}_\text{matt}.
\label{tilde-H}
\end{eqnarray}
In eq.~\eqref{tilde-H}, the DMP expansion parameter $\epsilon$ is defined in eq.~\eqref{epsilon-Dm2-ren-def} together with $\Delta m^2_{ \text{ren} }$. $\Delta_{a}$ and $\Delta_{b}$ are defined in eq.~\eqref{ratios-def}. In the DMP-UV perturbation theory we use $\epsilon$ and the six $\widetilde{\alpha}_{\beta \gamma}$ as the expansion parameters. 

In the second line in the right-hand side of eq.~\eqref{tilde-H}, $\widetilde{H}_\text{vac}$, can de decomposed into the zeroth order and first order terms. We regard the first two terms in $\widetilde{H}_\text{vac}$ as $\widetilde{H}_\text{vac}^{(0)}$~\cite{Minakata:2015gra}, and the third term as $\widetilde{H}_\text{vac}^{(1)}$. The third line in eq.~\eqref{tilde-H}, $\widetilde{H}_\text{matt}$ can be written as $\widetilde{H}_\text{matt} = \widetilde{H}_{ \text{matt} }^{(0)} + \widetilde{H}_\text{ UV }^{(1)} + \widetilde{H}_\text{ UV }^{(2)}$ where 
\begin{eqnarray}
&& \widetilde{H}_{ \text{matt} }^{(0)} = 
\left[
\begin{array}{ccc}
\Delta_{a} - \Delta_{b} & 0 & 0 \\
0 & - \Delta_{b} & 0 \\
0 & 0 & - \Delta_{b} \\
\end{array}
\right], 
\label{tilde-H-matt-0th} 
\end{eqnarray}
and 
\begin{eqnarray}
\widetilde{H}_\text{ UV }^{(1)} 
&=& 
\Delta_{b} 
U_{23}^{\dagger} 
\left[ 
\begin{array}{ccc}
2 \widetilde{\alpha}_{ee} \left( 1 - \frac{ \Delta_{a} }{ \Delta_{b} } \right) & \widetilde{\alpha}_{\mu e}^* & \widetilde{\alpha}_{\tau e}^* \\
\widetilde{\alpha}_{\mu e} & 2 \widetilde{\alpha}_{\mu \mu}  & \widetilde{\alpha}_{\tau \mu}^* \\
\widetilde{\alpha}_{\tau e}  & \widetilde{\alpha}_{\tau \mu} & 2 \widetilde{\alpha}_{\tau \tau} \\
\end{array}
\right] 
U_{23} 
\equiv 
\Delta_{b} F, 
\nonumber \\ 
\widetilde{H}_\text{ UV }^{(2)} 
&=& 
- \Delta_{b} U_{23}^{\dagger} 
\left[
\begin{array}{ccc}
\widetilde{\alpha}_{ee}^2 \left( 1 - \frac{ \Delta_{a} }{ \Delta_{b} } \right) + |\widetilde{\alpha}_{\mu e}|^2 + |\widetilde{\alpha}_{\tau e}|^2 & 
\widetilde{\alpha}_{\mu e}^* \widetilde{\alpha}_{\mu \mu} + \widetilde{\alpha}_{\tau e}^* \widetilde{\alpha}_{\tau \mu} & 
\widetilde{\alpha}_{\tau e}^* \widetilde{\alpha}_{\tau \tau} \\
\widetilde{\alpha}_{\mu e} \widetilde{\alpha}_{\mu \mu} + \widetilde{\alpha}_{\tau e} \widetilde{\alpha}_{\tau \mu}^* & 
\widetilde{\alpha}_{\mu \mu}^2 + |\widetilde{\alpha}_{\tau \mu}|^2 & 
\widetilde{\alpha}_{\tau \mu}^* \widetilde{\alpha}_{\tau \tau} \\
\widetilde{\alpha}_{\tau e} \widetilde{\alpha}_{\tau \tau} & 
\widetilde{\alpha}_{\tau \mu} \widetilde{\alpha}_{\tau \tau} & 
\widetilde{\alpha}_{\tau \tau}^2 \\
\end{array}
\right] 
U_{23}. 
\nonumber \\ 
\label{tilde-H-matt-1st} 
\end{eqnarray}
We have defined the $F$ matrix in the first line in eq.~\eqref{tilde-H-matt-1st}. In the same line, we define the matrix removing the 2-3 rotation from $F$, just full of the $\alpha$ parameters as $A \equiv U_{23} F U_{23}^{\dagger}$. It is for convenience for usage in section~\ref{sec:remarks-alpha}. 

\subsection{Unperturbed and perturbed Hamiltonian in the tilde basis}
\label{sec:unper-perturbed-H}

To formulate the DMP-UV perturbation theory, we decompose the tilde basis Hamiltonian into the zeroth and first order terms as $\widetilde{H} = \widetilde{H}^{(0)} + \widetilde{H}^{(1)}$. The unperturbed (zeroth-order) Hamiltonian is given by $\widetilde{H}^{(0)}=\widetilde{H}_{ \text{vac} }^{(0)} + \widetilde{H}_\text{matt}^{(0)}$. We make a phase redefinition 
\begin{eqnarray}
\widetilde{\nu} = \exp{ [ i \int^{x} dx^{\prime} \Delta_{b} (x^{\prime}) ] } \widetilde{\nu}^{\prime},
\end{eqnarray}
which is valid even for non-uniform matter density, to get rid of the NC potential term from $\widetilde{H}_{ \text{matt} }^{(0)}$. Then, the unperturbed part of the Hamiltonian 
$( \widetilde{H}^{(0)} )^{\prime}$ is given by 
\begin{eqnarray} 
( \widetilde{H}^{(0)} )^{\prime} &=& 
\Delta_{ \text{ren} } 
\left\{
\left[
\begin{array}{ccc}
\frac{ a(x) }{ \Delta m^2_{ \text{ren} }}  + s^2_{13} & 0 & c_{13} s_{13} \\
0 & 0 & 0 \\
c_{13} s_{13}  & 0 & c^2_{13} 
\end{array}
\right] 
+
\epsilon 
\left[
\begin{array}{ccc}
s^2_{12} & 0 & 0 \\
0 & c^2_{12} & 0 \\
0 & 0 & s^2_{12} 
\end{array}
\right]  \right\}. 
\label{tilde-H0}
\end{eqnarray}
Hereafter, we omit the prime symbol and use eq.~\eqref{tilde-H0} as the unperturbed part of the Hamiltonian. This is nothing but the zeroth order Hamiltonian used in ref.~\cite{Minakata:2015gra}. 

The perturbed Hamiltonian is then given by
\begin{eqnarray} 
\widetilde{H}^{(1)} = \widetilde{H}_{ \text{vac} }^{(1)} + \widetilde{H}_\text{ UV }^{(1)} + \widetilde{H}_\text{ UV }^{(2)}
\label{tilde-H1}
\end{eqnarray}
where $\widetilde{H}_{ \text{vac} }^{(1)}$ is the third term in the first line in eq.~\eqref{tilde-H}, and $\widetilde{H}_\text{ UV }^{(1)}$ and $\widetilde{H}_\text{ UV }^{(2)}$ are defined in eq.~\eqref{tilde-H-matt-1st}. In the following computation, we drop the second-order term (the last term) in eq.~\eqref{tilde-H1} because we confine ourselves into the zeroth and first order terms in the $\nu$SM and the UV parameters in this paper. 

\subsection{$U_{13}(\phi)$ rotation to the hat basis}

We diagonalize the 1-3 sector of $\widetilde{H}^{(0)}$ by doing the $U_{13}(\phi)$ rotation, where 
\begin{eqnarray} 
&& 
U_{13}(\phi) 
= 
\left[
\begin{array}{ccc}
c_{\phi}  & 0 &  s_{\phi} \\
0 & 1 & 0 \\
- s_{\phi} & 0 & c_{\phi}  \\
\end{array}
\right], 
\hspace{8mm}
U_{13}(\phi)^{\dagger}
= 
\left[
\begin{array}{ccc}
c_{\phi}  & 0 & - s_{\phi} \\
0 & 1 & 0 \\
s_{\phi} & 0 & c_{\phi}  \\
\end{array}
\right].
\label{U13-def}
\end{eqnarray}
After this rotation the neutrino basis becomes the hat basis, 
\begin{equation}
\ket{\hat{\nu}}=U^\dagger_{13}(\phi)\ket{\widetilde{\nu}}=U^\dagger_{13}(\phi)U^\dagger_{23}(\theta_{23})\ket{\nu}\,,
\end{equation}
and the $\nu$SM part of the Hamiltonian is given in the SOL convention by 
\begin{eqnarray}
&&
\hat{H} _\text{$\nu$SM}
= U^\dagger_{13}(\phi)~\widetilde{H}_\text{$\nu$SM}~U_{13}(\phi) 
\nonumber \\
&=& 
\frac{1}{2E}
\left[
\begin{array}{ccc}
\lambda_{-} & 0 & 0 \\
0 & \lambda_{0} & 0 \\
0 & 0 & \lambda_{+} 
\end{array}
\right] 
+ \epsilon c_{12} s_{12} \Delta_{ \text{ren} }
\left[
\begin{array}{ccc}
0 & 
\cos ( \phi - \theta_{13} ) e^{ i \delta} & 
0 \\
\cos ( \phi - \theta_{13} ) e^{ - i \delta} & 
0 & 
\sin ( \phi - \theta_{13} ) e^{ - i \delta} \\
0 & 
\sin ( \phi - \theta_{13} ) e^{ i \delta} & 
0 \\
\end{array}
\right]. 
\nonumber \\
\label{hat-H}
\end{eqnarray}
In eq.~\eqref{hat-H}, the first term $\hat{H}^{(0)}$ is the unperturbed term with the eigenvalues 
\begin{eqnarray} 
\lambda_{-} &=& 
\frac{ 1 }{ 2 } \left[
\left( \Delta m^2_{ \text{ren} } + a \right) - {\rm sign}(\Delta m^2_{ \text{ren} }) \sqrt{ \left( \Delta m^2_{ \text{ren} } - a \right)^2 + 4 s^2_{13} a \Delta m^2_{ \text{ren} } }
\right] 
+ \epsilon \Delta m^2_{ \text{ren} } s^2_{12},
\nonumber\\
\lambda_{0} &=&  c^2_{12} ~\epsilon ~\Delta m^2_{ \text{ren} },
\label{lambda-pm0}  
\\
\lambda_{+} &=& 
\frac{ 1 }{ 2 } \left[
\left( \Delta m^2_{ \text{ren} } + a \right) + {\rm sign}(\Delta m^2_{ \text{ren} }) \sqrt{ \left( \Delta m^2_{ \text{ren} } - a \right)^2 + 4 s^2_{13} a \Delta m^2_{ \text{ren} } }
\right] 
+ \epsilon \Delta m^2_{ \text{ren} } s^2_{12}, 
\nonumber 
\end{eqnarray}
and the second term is the first order perturbation. 
The diagonalization determines $\phi$ as 
\begin{eqnarray} 
\cos 2 \phi &=& 
\frac{ \Delta m^2_{ \text{ren} } \cos 2\theta_{13} - a }{ \lambda_{+} - \lambda_{-} } 
= \frac{ \cos 2\theta_{13} - r_{a} }{ 
\sqrt{ 1 + r_{a}^2 - 2 r_{a} \cos 2\theta_{13} } }, 
\nonumber \\
\sin 2 \phi &=& \frac{ \Delta m^2_{ \text{ren} } \sin 2\theta_{13} }{ \lambda_{+} - \lambda_{-} } 
= \frac{ \sin 2\theta_{13} }{ 
\sqrt{ 1 + r_{a}^2 - 2 r_{a} \cos 2\theta_{13} } }. 
\label{2phi}
\end{eqnarray}
So far, it is the $\nu$SM treatment given in ref.~\cite{Minakata:2015gra}. 

The UV part of the Hamiltonian is given
\begin{eqnarray} 
&& 
\hat{H}^{(1)}_\text{UV} = 
U^\dagger_{13}(\phi) 
\widetilde{H}_\text{ UV }^{(1)} 
U_{13}(\phi) 
= 
U^\dagger_{13}(\phi) 
\Delta_{b} F 
U_{13}(\phi) 
\equiv 
\Delta_{b} K 
\label{K-def}
\end{eqnarray}
where we have defined the $K$ matrix as 
\begin{eqnarray} 
&& 
K \equiv 
U^\dagger_{13}(\phi) F U_{13}(\phi) 
= 
\left[
\begin{array}{ccc}
K_{11} & K_{12} & K_{13} \\
K_{21} & K_{22} & K_{23} \\
K_{31} & K_{32} & K_{33} \\
\end{array}
\right] 
\nonumber \\
&=&
\left[
\begin{array}{ccc}
c^2_{\phi} F_{11} + s^2_{\phi} F_{33} - c_{\phi} s_{\phi} \left( F_{13} + F_{31} \right) & 
c_{\phi} F_{12} - s_{\phi} F_{32} & 
c^2_{\phi} F_{13} - s^2_{\phi} F_{31} + c_{\phi} s_{\phi} \left( F_{11} - F_{33} \right) \\
c_{\phi} F_{21} - s_{\phi} F_{23} & 
F_{22} & 
s_{\phi} F_{21} + c_{\phi} F_{23} \\
c^2_{\phi} F_{31} - s^2_{\phi} F_{13} + c_{\phi} s_{\phi} \left( F_{11} - F_{33} \right) & 
s_{\phi} F_{12} + c_{\phi} F_{32} & 
s^2_{\phi} F_{11} + c^2_{\phi} F_{33} + c_{\phi} s_{\phi} \left( F_{13} + F_{31} \right) \\
\end{array}
\right]. 
\nonumber \\
\label{Kij-by-Fij}
\end{eqnarray}
The $F$ matrix is defined in eq.~\eqref{tilde-H-matt-1st}. The explicit expressions of the elements $K_{ij}$ are given in eq.~\eqref{Kij-elements}. 

\subsection{$U_{12}(\psi)$ rotation to the bar (energy eigenstate) basis}

Since $\lambda_-$ and $\lambda_0$ cross at the solar resonance, $a \approx \epsilon \Dmsqren \cos 2 \theta_{12}/\cos^2 \theta_{13}$, to describe the physics near this degeneracy we need to diagonalize the (1-2) submatrix of $\hat{H} _\text{$\nu$SM}$ using $U_{12}(\psi)$: 
\begin{equation}
U_{12} (\psi) = 
\left[
\begin{array}{ccc}
c_\psi & s_\psi e^{ i \delta} & 0 \\
- s_\psi e^{ - i \delta} & c_\psi & 0 \\
0 & 0 & 1
\end{array}
\right], 
\hspace{8mm}
U_{12} (\psi)^{\dagger} 
= \left[
\begin{array}{ccc}
c_\psi & - s_\psi e^{ i \delta} & 0 \\
s_\psi e^{ - i \delta} & c_\psi & 0 \\
0 & 0 & 1
\end{array}
\right].
\label{U12-def}
\end{equation}
The new neutrino basis is 
\begin{equation}
\ket{\bar{\nu}} = U^\dagger_{12}(\psi)\ket{\hat{\nu}} = U^\dagger_{12}(\psi, \delta) U^\dagger_{13}(\phi)U^\dagger_{23}(\theta_{23})\ket{\nu}, 
\end{equation}
and the $\nu$SM part of the Hamiltonian 
\begin{eqnarray} 
&&
\bar{H} _\text{$\nu$SM} 
= 
U_{12} (\psi)^{\dagger} \hat{H}_\text{$\nu$SM} U_{12} (\psi) 
= 
\frac{1}{2E}
\left[
\begin{array}{ccc}
\lambda_{1} & 0 & - s_\psi A_S \\
0 & \lambda_{2} & c_\psi A_S e^{ - i \delta} \\ 
- s_\psi A_S & c_\psi A_S e^{ i \delta} & \lambda_{3}
\end{array}
\right], 
\label{barH-SM} 
\end{eqnarray}
where $A_C$ and $A_S$ characterize the first order correction effects, and are given by 
\begin{eqnarray} 
&&
A_C \equiv \epsilon c_{12} s_{12} c_{\phi - \theta_{13}} \Dmsqren,
\nonumber \\
&&
A_S \equiv \epsilon c_{12} s_{12} s_{\phi - \theta_{13}} \Dmsqren.
\end{eqnarray}
The eigenvalues are given by $\lambda_{3} = \lambda_{+}$, and 
\begin{eqnarray} 
&&
\lambda_{1} = 
\frac{1}{2E} 
\left( c^2_\psi \lambda_{-} + s^2_\psi \lambda_{0} - 2 c_\psi s_\psi \epsilon c_{12} s_{12} c_{\phi - \theta_{13}} \Dmsqren \right),
\nonumber \\
&&
\lambda_{2} = 
\frac{1}{2E} 
\left( s^2_\psi \lambda_{-} + c^2_\psi \lambda_{0} + 2 c_\psi s_\psi \epsilon c_{12} s_{12} c_{\phi - \theta_{13}} \Dmsqren \right).
\label{lambda-12}
\end{eqnarray}

Finally, the UV part of the first order bar-basis Hamiltonian is given by 
\begin{eqnarray} 
&& 
\bar{H}_{UV} 
= U_{12} (\psi)^{\dagger} \hat{H}_{UV}^{(1)} U_{12} (\psi) 
= 
\Delta_{b}
U_{12} (\psi)^{\dagger} K U_{12} (\psi) 
\equiv 
\Delta_{b} G, 
\label{barH-UV}
\end{eqnarray}
where $\Delta_{b} \equiv \frac{b}{2E}$ and we have defined the $G$ matrix 
\begin{eqnarray} 
&&
G \equiv U_{12} (\psi)^{\dagger} K U_{12} (\psi) 
= 
\left[
\begin{array}{ccc}
G_{11} & G_{12} & G_{13} \\
G_{21} & G_{22} & G_{23} \\
G_{31} & G_{32} & G_{33} \\
\end{array}
\right].
\label{G-def}
\end{eqnarray}
The expressions of the $G$ matrix elements in terms of the $K_{ij}$'s are given by 
\begin{eqnarray} 
&& G_{11} 
=
c^2_\psi K_{11} + s^2_\psi K_{22} 
- c_\psi s_\psi \left( e^{ i \delta} K_{21} + e^{ - i \delta} K_{12} \right), 
\nonumber \\
&& G_{12} 
=
e^{ i \delta} \left[ c_\psi s_\psi \left( K_{11} - K_{22} \right) 
+ \left( c^2_\psi e^{ - i \delta} K_{12} - s^2_\psi e^{ i \delta} K_{21} \right) \right] 
= \left( G_{21} \right)^*, 
\nonumber \\
&& G_{22} 
= 
s^2_\psi K_{11} + c^2_\psi K_{22} 
+ c_\psi s_\psi \left( e^{ i \delta} K_{21} + e^{ - i \delta} K_{12} \right), 
\nonumber \\
&& G_{13} 
= 
c_\psi K_{13} - s_\psi e^{ i \delta} K_{23} 
= \left( G_{31} \right)^*, 
\nonumber \\
&& G_{23} 
= 
e^{ - i \delta} \left( s_\psi K_{13} + c_\psi e^{ i \delta} K_{23} \right) 
= \left( G_{32} \right)^*, 
\nonumber \\
&& G_{33} 
= 
K_{33}. 
\label{Gij-elements}
\end{eqnarray} 
Notice that $K_{ji} = K_{ij}^*$ and $K_{ii}$ are real.  

\subsubsection{The bar-basis Hamiltonian in the SOL convention: Summary}

To summarize, the unperturbed and perturbed parts of the bar-basis Hamiltonian can be written as $\bar{H} = \bar{H}^{(0)} + \bar{H}^{(1)}$, where  
\begin{eqnarray} 
&& 
\bar{H}^{(0)} 
= 
\frac{1}{2E} 
\left[
\begin{array}{ccc}
\lambda_1 & 0 & 0 \\
0 & \lambda_2 & 0 \\
0 & 0 & \lambda_3 
\end{array}
\right], 
\nonumber \\
&& 
\bar{H}^{(1)}
= 
\epsilon c_{12} s_{12} s_{( \phi - \theta_{13} )} \Delta_{ \text{ren} } 
\left[
\begin{array}{ccc}
0 & 0 & - s_\psi \\
0 & 0 & c_\psi e^{ - i \delta} \\
- s_\psi & c_\psi e^{ i \delta} & 0 
\end{array}
\right] 
+ \Delta_{b} G, 
\label{barH-summary}
\end{eqnarray}
where $\Delta_{ \text{ren} } \equiv \frac{\Dmsqren}{2E}$ is defined in eq.~\eqref{ratios-def}.

\subsection{Calculation of the bar-basis $\bar{S}$ matrix}

To calculate $\bar{S}$ matrix we define $\Omega(x)$ as
\begin{eqnarray} 
\Omega(x) = e^{i \bar{H}^{(0)} x} \bar{S} (x).
\label{def-omega}
\end{eqnarray}
Using $i \frac{d}{dx} \bar{S} = \bar{H} (x) \bar{S}$, $\Omega(x)$ obeys the evolution equation
\begin{eqnarray} 
i \frac{d}{dx} \Omega(x) = H_{1} \Omega(x) 
\label{omega-evolution}
\end{eqnarray}
where
\begin{eqnarray} 
H_{1} \equiv e^{i \bar{H}^{(0)} x} \bar{H}^{(1)} e^{-i \bar{H}^{(0)} x} .
\label{def-H1}
\end{eqnarray}
Then, $\Omega(x)$ can be computed perturbatively as
\begin{eqnarray} 
\Omega(x) &=& 1 + 
(-i) \int^{x}_{0} dx' H_{1} (x') + 
(-i)^2 \int^{x}_{0} dx' H_{1} (x') \int^{x'}_{0} dx'' H_{1} (x'') 
+ \cdot \cdot \cdot,
\label{Omega-expansion}
\end{eqnarray}
and the $\bar{S}$ matrix is given by
\begin{eqnarray} 
\bar{S} (x) =  
e^{-i \bar{H}^{(0)} x} \Omega(x). 
\label{hat-Smatrix}
\end{eqnarray}
For simplicity of the expressions we use the notation 
\begin{eqnarray} 
h_{i} = \frac{ \lambda_{i} }{2E}
\end{eqnarray}
which leads to 
\begin{eqnarray} 
e^{ \pm i \bar{H}^{(0)} x } &=& 
\left[
\begin{array}{ccc}
e^{ \pm i h_{1} x } & 0 & 0 \\
0 & e^{ \pm i h_{2} x } & 0 \\
0 & 0 & e^{ \pm i h_{3} x } \\
\end{array}
\right]. 
\end{eqnarray}
$H_{1}$ can be given as the sum of $\nu$SM and UV terms, 
$H_{1} = H_{1}^{\nu \text{SM}} + H_{1}^{\text{UV}}$, where 
\begin{eqnarray} 
&& 
H_{1}^{\nu \text{SM}} 
=
\widetilde{\epsilon} \Delta_{ \text{ren} } 
\left[
\begin{array}{ccc}
0 & 0 & - s_{\psi} e^{ - i ( h_{3} - h_{1} ) x } \\
0 & 0 & c_{\psi} e^{ - i \delta} e^{ - i ( h_{3} - h_{2} ) x } \\
- s_{\psi} e^{ i ( h_{3} - h_{1} ) x } & c_{\psi} e^{ i \delta} e^{ i ( h_{3} - h_{2} ) x } & 0 \\
\end{array}
\right], 
\nonumber \\
&& 
H_{1}^{UV}
= 
\Delta_{b} 
\left[
\begin{array}{ccc}
G_{11} & e^{ - i ( h_{2} - h_{1} ) x } G_{12} & e^{ - i ( h_{3} - h_{1} ) x } G_{13} \\
e^{ i ( h_{2} - h_{1} ) x } G_{21} & G_{22} & e^{ - i ( h_{3} - h_{2} ) x } G_{23} \\
e^{ i ( h_{3} - h_{1} ) x } G_{31} & e^{ i ( h_{3} - h_{2} ) x } G_{32} & G_{33} \\
\end{array}
\right], 
\nonumber 
\end{eqnarray}
where we have defined $\widetilde{\epsilon}$:
\begin{eqnarray} 
&&
\widetilde{\epsilon} \equiv \epsilon c_{12} s_{12} s_{ (\phi - \theta_{13}) }. 
\label{widetilde-epsilon-def} 
\end{eqnarray}
Then, the first order terms in $\Omega(x)$ and subsequently $\bar{S}^{(1)}$ can be obtained as 
\begin{eqnarray} 
&& 
\hspace{-12mm}
\bar{S}^{(1)} = 
\widetilde{\epsilon} 
\left[
\begin{array}{ccc}
0 & 0 & 
- s_{\psi} \frac{ \Delta_{ \text{ren} } }{ h_{3} - h_{1} } 
\left\{ e^{ - i h_{3} x } - e^{ - i h_{1} x } \right\} \\
0 & 0 & 
c_{\psi} e^{ - i \delta} \frac{ \Delta_{ \text{ren} } }{ h_{3} - h_{2} } 
\left\{ e^{ - i h_{3} x } - e^{ - i h_{2} x } \right\} \\
- s_{\psi} \frac{ \Delta_{ \text{ren} } }{ h_{3} - h_{1} } 
\left\{ e^{ - i h_{3} x } - e^{ - i h_{1} x }  \right\} & 
c_{\psi} e^{ i \delta} \frac{ \Delta_{ \text{ren} } }{ h_{3} - h_{2} } 
\left\{ e^{ - i h_{3} x } - e^{ - i h_{2} x } \right\} & 0 \\
\end{array}
\right]
\nonumber \\
&+& 
\left[
\begin{array}{ccc}
G_{11} ( - i \Delta_{b} x ) e^{ - i h_{1} x } & 
G_{12} \frac{ \Delta_{b} }{ h_{2} - h_{1} } \left\{ e^{ - i h_{2} x } - e^{ - i h_{1} x } \right\} & 
G_{13} \frac{ \Delta_{b} }{ h_{3} - h_{1} } \left\{ e^{ - i h_{3} x } - e^{ - i h_{1} x } \right\} \\
G_{21} \frac{ \Delta_{b} }{ h_{2} - h_{1} } \left\{ e^{ - i h_{2} x } - e^{ - i h_{1} x } \right\} & 
G_{22} ( - i \Delta_{b} x ) e^{ - i h_{2} x } & 
G_{23} \frac{ \Delta_{b} }{ h_{3} - h_{2} } \left\{ e^{ - i h_{3} x } - e^{ - i h_{2} x } \right\} \\
G_{31} \frac{ \Delta_{b} }{ h_{3} - h_{1} } \left\{ e^{ - i h_{3} x } - e^{ - i h_{1} x } \right\} & 
G_{32} \frac{ \Delta_{b} }{ h_{3} - h_{2} } \left\{ e^{ - i h_{3} x } - e^{ - i h_{2} x } \right\} & 
G_{33} ( - i \Delta_{b}x ) e^{ - i h_{3} x } \\
\end{array}
\right]. ~~
%
\label{checkS-elements}
\end{eqnarray}

To summarize: 
The bar-basis $S$ matrix to first order in $\epsilon$ and $\widetilde{ \alpha }_{\beta \gamma}$ is given by 
\begin{eqnarray} 
&& 
\bar{S} 
= 
\left[
\begin{array}{ccc}
e^{ - i h_{1} x } & 0 & 0 \\
0 & e^{ - i h_{2} x } & 0 \\
0 & 0 & e^{ - i h_{3} x } \\
\end{array}
\right] 
+
\left[
\begin{array}{ccc}
\bar{S}^{(1)}_{11} & \bar{S}^{(1)}_{12} & \bar{S}^{(1)}_{13} \\
\bar{S}^{(1)}_{21} & \bar{S}^{(1)}_{22} & \bar{S}^{(1)}_{23} \\
\bar{S}^{(1)}_{31} & \bar{S}^{(1)}_{32} & \bar{S}^{(1)}_{33} \\
\end{array}
\right], 
\label{barS(1)-def} 
\end{eqnarray}
where $\bar{S}^{(1)}$ matrix elements are given in eq.~\eqref{checkS-elements}. By using the expressions of the $G$ matrix elements in eq.~\eqref{Gij-elements}, they can be written by using the $K$ matrix elements. The resulting expressions of $\bar{S}^{(1)}$ elements in first order in the UV $\widetilde{\alpha}$ parameters are as follows: 
\begin{eqnarray} 
&&
\bar{S}^{(1)} _{11} 
= 
\left[ 
c^2_\psi K_{11} + s^2_\psi K_{22} 
- c_\psi s_\psi \left( e^{ i \delta} K_{21} + e^{ - i \delta} K_{12} \right) 
\right]
( - i \Delta_{b} x ) e^{ - i h_{1} x },  
\nonumber \\
&& 
\bar{S}^{(1)} _{22} 
= 
\left[
s^2_\psi K_{11} + c^2_\psi K_{22} 
+ c_\psi s_\psi \left( e^{ i \delta} K_{21} + e^{ - i \delta} K_{12} \right) 
\right]
( - i \Delta_{b} x ) e^{ - i h_{2} x }, 
\nonumber \\
&& 
\bar{S}^{(1)} _{33} 
= 
K_{33} 
( - i \Delta_{b} x ) e^{ - i h_{3} x }, 
\nonumber \\
&& 
\bar{S}^{(1)} _{12} 
=
e^{ i \delta} \left[ c_\psi s_\psi \left( K_{11} - K_{22} \right) 
+ \left( c^2_\psi e^{ - i \delta} K_{12} - s^2_\psi e^{ i \delta} K_{21} \right) \right] 
\frac{ \Delta_{b} }{ h_{2} - h_{1} } \left\{ e^{ - i h_{2} x } - e^{ - i h_{1} x } \right\}, 
\nonumber \\
&& 
\bar{S}^{(1)} _{21} 
=
e^{ - i \delta} \left[ c_\psi s_\psi \left( K_{11} - K_{22} \right) 
+ \left( c^2_\psi e^{ i \delta} K_{21} - s^2_\psi e^{ - i \delta} K_{12} \right) \right] 
\frac{ \Delta_{b} }{ h_{2} - h_{1} } \left\{ e^{ - i h_{2} x } - e^{ - i h_{1} x } \right\}, 
\nonumber \\
&&
\bar{S}^{(1)} _{13} 
=
\left[ 
\left( c_\psi K_{13} - s_\psi e^{ i \delta} K_{23} \right) 
\Delta_{b} - \widetilde{\epsilon} s_{\psi} \Delta_{ \text{ren} } \right] 
\frac{ 1 }{ h_{3} - h_{1} } \left\{ e^{ - i h_{3} x } - e^{ - i h_{1} x } \right\}, 
\nonumber \\
&& 
\bar{S}^{(1)} _{31} 
=
\left[ 
\left( c_\psi K_{31} - s_\psi e^{ - i \delta} K_{32} \right) 
\Delta_{b} - \widetilde{\epsilon} s_{\psi} \Delta_{ \text{ren} } \right] 
\frac{ 1 }{ h_{3} - h_{1} } \left\{ e^{ - i h_{3} x } - e^{ - i h_{1} x } \right\}, 
\nonumber \\
&& 
\bar{S}^{(1)} _{23} 
= 
e^{ - i \delta} 
\left[ 
\left( s_\psi K_{13} + c_\psi e^{ i \delta} K_{23} \right) 
\Delta_{b} + \widetilde{\epsilon} c_{\psi} \Delta_{ \text{ren} } \right] 
\frac{ 1 }{ h_{3} - h_{2} } \left\{ e^{ - i h_{3} x } - e^{ - i h_{2} x } \right\}, 
\nonumber \\
&& 
\bar{S}^{(1)} _{32} 
= 
e^{ i \delta} 
\left[ 
\left( s_\psi K_{31} + c_\psi e^{ - i \delta} K_{32} \right) 
\Delta_{b} + \widetilde{\epsilon} c_{\psi} \Delta_{ \text{ren} } \right] 
\frac{ 1 }{ h_{3} - h_{2} } \left\{ e^{ - i h_{3} x } - e^{ - i h_{2} x } \right\}, 
\label{barS(1)-expressions}
\end{eqnarray}

\subsection{The relations between various bases} 

A clarifying note on the relations between the various bases may help. Using the definitions of the tilde and hat bases 
\begin{eqnarray} 
&& \widetilde{H} = 
( U_{13} U_{12} ) \check{H} ( U_{13} U_{12} )^{\dagger}, 
\hspace{10mm} 
\hat{H} = U^\dagger_{13}(\phi) \widetilde{H} U_{13}(\phi) 
\nonumber 
\end{eqnarray}
the relationship between the bar basis Hamiltonian $\bar{H}$ and the vacuum mass eigenstate check basis Hamiltonian $\check{H}$ is given by 
\begin{eqnarray} 
&& 
\bar{H} 
= U_{12} (\psi)^{\dagger} \hat{H} U_{12} (\psi) 
= 
U_{12}^{\dagger} (\psi) U^\dagger_{13}(\phi) 
U_{13} U_{12} \check{H} U_{12}^{\dagger} U_{13}^{\dagger}
U_{13}(\phi) U_{12} (\psi), 
\nonumber \\
&& 
\check{H} 
= 
U_{12}^{\dagger} U_{13}^{\dagger} U_{13}(\phi) U_{12} (\psi) 
\bar{H} 
U_{12}^{\dagger} (\psi) U^\dagger_{13}(\phi) U_{13} U_{12}. 
\label{bar-check}
\end{eqnarray}
The bar $\bar{H}$ basis is the basis in which $H_{0}$ is diagonalized, and therefore the perturbation theory is formulated by using the bar basis. 
In eq.~\eqref{bar-check} and hereafter, $U_{12}$, $U_{13}$ etc. with no specified arguments assumes the vacuum angles $\theta_{12}$ and $\theta_{13}$, respectively, as the arguments.

In our discussion with UV, the non-unitary transformation is involved in the relation between the flavor and the vacuum mass eigenstate (check) basis as 
\begin{eqnarray} 
\nu_{\alpha} = N_{\alpha i} \check{\nu}_{i} 
= \left\{ ( 1 - \alpha ) U \right\}_{\alpha i} \check{\nu}_{i}. 
\label{flavor-check}
\end{eqnarray}
Then, the relationship between the flavor basis Hamiltonian $H_{ \text{flavor} }$ and the bar basis one $\bar{H}$ becomes 
\begin{eqnarray}
H_{ \text{flavor} } 
&=& 
\left\{ ( 1 - \alpha ) U \right\} \check{H} \left\{ ( 1 - \alpha ) U \right\}^{\dagger} 
\nonumber \\
&=& 
( 1 - \alpha ) U_{23} 
U_{13}(\phi) U_{12} (\psi) 
\bar{H} 
U_{12}^{\dagger} (\psi) U^\dagger_{13}(\phi) 
U_{23}^{\dagger} ( 1 - \alpha )^{\dagger}.
\label{H-flavor-bar}
\end{eqnarray}
Then, the flavor basis $S$ matrix is related to $\bar{S}$ matrix as
\begin{eqnarray}
S_{ \text{flavor} } 
&=& 
( 1 - \alpha ) U_{23} 
U_{13}(\phi) U_{12} (\psi) 
\bar{S} 
U_{12}^{\dagger} (\psi) U^\dagger_{13}(\phi) 
U_{23}^{\dagger} ( 1 - \alpha )^{\dagger}.
\label{S-flavor-bar}
\end{eqnarray}

Notice that the structure of the $S_{ \text{flavor} }$ matrix in eq.~\eqref{S-flavor-bar} is very similar to the one in the solar-resonance perturbation theory, eq.~(51) in ref.~\cite{Martinez-Soler:2019noy}. That is why we will see the similar expressions of the various quantities as functions of $K$ and $G$ matrices. But, we must note that the similarity is superficial because the energy eigenstate bases in matter by which the perturbation theory is formulated, the bar basis here and the hat basis in ref.~\cite{Martinez-Soler:2019noy}, is different from each other. Using the notation $\varphi$ for $\theta_{12}$ in matter, it is given by $\hat{H} = U_{12}^{\dagger} (\varphi) U_{12} \check{H} U_{12}^{\dagger} U_{12} (\varphi)$ in ref.~\cite{Martinez-Soler:2019noy}, which is very different from eq.~\eqref{bar-check} in our case. 

\subsection{Calculation of the flavor basis $S$ matrix}
\label{sec:flavor-S}

Given the expression of the bar-basis $\bar{S}$ matrix in the zeroth and the first orders in eqs.~\eqref{barS(1)-def} and \eqref{barS(1)-expressions}, respectively, it is straightforward to compute the flavor basis $S$ matrix elements using eq.~\eqref{S-flavor-bar}. 

First, we calculate the first order terms with use of $\bar{S}^{(1)}$ matrix. In this calculation we can disregard the $( 1 - \alpha )$ and $( 1 - \alpha )^{\dagger}$ factors because we are interested in up to first-order of the $S_{ \text{flavor} }$ matrix. Then, what we should do is the three rotations in the 1-2, 1-3, and 2-3 spaces as in eq.~\eqref{S-flavor-bar}. For possible convenience of the readers we give an intermediate step, the tilde basis $\widetilde{S}^{(1)}$ matrix, 
\begin{eqnarray} 
&&
\widetilde{S}^{(1)} = U_{13}(\phi) \hat{S}^{(1)} U^\dagger_{13}(\phi)
=
U_{13}(\phi) U_{12} (\psi) 
\bar{S} 
U_{12}^{\dagger} (\psi) U^\dagger_{13}(\phi). 
\label{tildeS-def}
\end{eqnarray}
The computed results of $\widetilde{S}^{(1)}$ matrix elements are given in Appendix~\ref{sec:tildeS-summary}. 

Using $\widetilde{S}^{(1)}$ matrix elements, the flavor basis $S^{(1)}$ matrix is obtained as 
\begin{eqnarray}
&& 
S^{(1)} = U_{23} \widetilde{S}^{(1)} U^\dagger_{23} 
\nonumber \\
&&
\hspace{-14mm}
= 
\left[
\begin{array}{ccc}
\widetilde{S}^{(1)}_{11} & 
c_{23} \widetilde{S}^{(1)}_{12} + s_{23} \widetilde{S}^{(1)}_{13} & 
- s_{23} \widetilde{S}^{(1)}_{12} + c_{23} \widetilde{S}^{(1)}_{13} \\
c_{23} \widetilde{S}^{(1)}_{21} + s_{23} \widetilde{S}^{(1)}_{31} & 
c^2_{23} \widetilde{S}^{(1)}_{22} + s^2_{23} \widetilde{S}^{(1)}_{33} 
+ c_{23} s_{23} \left( \widetilde{S}^{(1)}_{32} + \widetilde{S}^{(1)}_{23} \right) & 
c^2_{23} \widetilde{S}^{(1)}_{23} - s^2_{23} \widetilde{S}^{(1)}_{32} 
+ c_{23} s_{23} \left( \widetilde{S}^{(1)}_{33} - \widetilde{S}^{(1)}_{22} \right) \\
- s_{23} \widetilde{S}^{(1)}_{21} + c_{23} \widetilde{S}^{(1)}_{31} & 
c^2_{23} \widetilde{S}^{(1)}_{32} - s^2_{23} \widetilde{S}^{(1)}_{23} 
+ c_{23} s_{23} \left( \widetilde{S}^{(1)}_{33} - \widetilde{S}^{(1)}_{22} \right) & 
s^2_{23} \widetilde{S}^{(1)}_{22} + c^2_{23} \widetilde{S}^{(1)}_{33} 
- c_{23} s_{23} \left( \widetilde{S}^{(1)}_{32} + \widetilde{S}^{(1)}_{23} \right) \\
\end{array}
\right]. 
\nonumber \\
\label{flavorS-by-tildeS}
\end{eqnarray}
Notice that the first-order $S^{(1)}$ matrix is necessary to obtain the EV part of the first-order probability as defined in eq.~\eqref{P-three-types}. For the UV part, the $S^{(0)}$ matrix elements suffice. 

For the zeroth-order flavor basis $S$ matrix $S^{(0)}$, we repeat the same calculation with use of the $\bar{S}^{(0)}$ matrix, the first term in eq.~\eqref{barS(1)-def}. The results are given below: 
\begin{eqnarray} 
S^{(0)}_{e e} 
&=& 
c^2_{\phi} \left( c^2_\psi e^{ - i h_{1} x } + s^2_\psi e^{ - i h_{2} x } \right) 
+ s^2_{\phi} e^{ - i h_{3} x }, 
\nonumber \\
S^{(0)}_{\mu \mu} 
&=&
c^2_{23} 
\left( s^2_\psi e^{ - i h_{1} x } + c^2_\psi e^{ - i h_{2} x } \right) 
+ s^2_{23} 
\left[ s^2_{\phi} \left( c^2_\psi e^{ - i h_{1} x } + s^2_\psi e^{ - i h_{2} x } \right) 
+ c^2_{\phi} e^{ - i h_{3} x } 
\right]
\nonumber \\
&-&
\sin 2\theta_{23} s_{\phi} c_\psi s_\psi \cos \delta 
\left( e^{ - i h_{2} x }  - e^{ - i h_{1} x } \right), 
\nonumber \\
S^{(0)}_{\tau \tau} 
&=&
s^2_{23} \left( s^2_\psi e^{ - i h_{1} x } + c^2_\psi e^{ - i h_{2} x } \right) 
+ c^2_{23} 
\left[ 
s^2_{\phi} \left( c^2_\psi e^{ - i h_{1} x } + s^2_\psi e^{ - i h_{2} x } \right) 
+ c^2_{\phi} e^{ - i h_{3} x } 
\right] 
\nonumber \\
&+&
\sin 2\theta_{23} s_{\phi} c_\psi s_\psi \cos \delta 
\left( e^{ - i h_{2} x }  - e^{ - i h_{1} x } \right), 
\nonumber \\
S^{(0)}_{e \mu} 
&=&
c_{23} c_{\phi} e^{ i \delta} 
c_\psi s_\psi \left( e^{ - i h_{2} x }  - e^{ - i h_{1} x } \right) 
+ s_{23} c_{\phi} s_{\phi} 
\left[ e^{ - i h_{3} x } - \left( c^2_\psi e^{ - i h_{1} x } + s^2_\psi e^{ - i h_{2} x } \right) 
\right],
\nonumber \\
S^{(0)}_{\mu e} 
&=& 
c_{23} c_{\phi} e^{ - i \delta} 
c_\psi s_\psi \left( e^{ - i h_{2} x }  - e^{ - i h_{1} x } \right) 
+ s_{23} c_{\phi} s_{\phi} 
\left[ e^{ - i h_{3} x } - \left( c^2_\psi e^{ - i h_{1} x } + s^2_\psi e^{ - i h_{2} x } \right) 
\right], 
\nonumber \\
S^{(0)}_{e \tau} 
&=&
- s_{23} c_{\phi} e^{ i \delta} 
c_\psi s_\psi \left( e^{ - i h_{2} x }  - e^{ - i h_{1} x } \right) 
+ c_{23} c_{\phi} s_{\phi} 
\left[ e^{ - i h_{3} x } - \left( c^2_\psi e^{ - i h_{1} x } + s^2_\psi e^{ - i h_{2} x } \right) 
\right], 
\nonumber \\
S^{(0)}_{\tau e} 
&=&
- s_{23} c_{\phi} e^{ - i \delta} 
c_\psi s_\psi \left( e^{ - i h_{2} x }  - e^{ - i h_{1} x }  \right) 
+ c_{23} c_{\phi} s_{\phi} 
\left[ e^{ - i h_{3} x } - \left( c^2_\psi e^{ - i h_{1} x } + s^2_\psi e^{ - i h_{2} x } \right) 
\right], 
\nonumber \\
S_{\mu \tau}^{(0)} 
&=& 
- \left( c^2_{23} e^{ - i \delta} - s^2_{23} e^{ i \delta} \right) 
s_{\phi} c_\psi s_\psi \left( e^{ - i h_{2} x }  - e^{ - i h_{1} x } \right) 
\nonumber \\
&+&
c_{23} s_{23} 
\left[ 
( s^2_{\phi} s^2_\psi - c^2_\psi ) e^{ - i h_{2} x } 
+ ( s^2_{\phi} c^2_\psi - s^2_\psi ) e^{ - i h_{1} x } 
+ c^2_{\phi} e^{ - i h_{3} x } 
\right],
\nonumber \\
S_{\tau \mu}^{(0)} 
&=&
- \left( c^2_{23} e^{ i \delta} - s^2_{23} e^{ - i \delta} \right) 
s_{\phi} c_\psi s_\psi \left( e^{ - i h_{2} x }  - e^{ - i h_{1} x } \right) 
\nonumber \\
&+&
c_{23} s_{23} 
\left[ 
( s^2_{\phi} s^2_\psi - c^2_\psi ) e^{ - i h_{2} x } 
+ ( s^2_{\phi} c^2_\psi - s^2_\psi ) e^{ - i h_{1} x } 
+ c^2_{\phi} e^{ - i h_{3} x } 
\right]. 
\label{flavorS0}
\end{eqnarray}
Notice that the generalized T invariance, $S^{(0)}_{ij} \vert_{i \leftrightarrow j} (c \rightarrow c^*) = S^{(0)}_{ji} (c)$  holds, where $c$ denotes the all complex numbers involved.

\section{Tilde basis $\widetilde{S}^{(1)}$ matrix: Summary} 
\label{sec:tildeS-summary}

The first-order tilde basis $\widetilde{S}^{(1)}$ matrix is defined in eq.~\eqref{tildeS-def} for the given expression of the bar-basis $\bar{S}^{(1)}$ matrix in eq.~\eqref{barS(1)-expressions}. The obtained results of $\widetilde{S}^{(1)}$ matrix elements are given below: 
\begin{eqnarray}
&& 
\widetilde{S}^{(1)}_{11} = 
( - i \Delta_{b} x ) 
\biggl\{
c^2_{\phi} K_{11} \left( s^4_\psi e^{ - i h_{2} x } + c^4_\psi e^{ - i h_{1} x } \right) 
+ c^2_{\phi} c^2_\psi s^2_\psi K_{22} \left( e^{ - i h_{2} x } + e^{ - i h_{1} x } \right) 
+ s^2_{\phi} K_{33} e^{ - i h_{3} x } 
\nonumber \\
&+& 
c^2_{\phi} c_\psi s_\psi 
\left( e^{ i \delta} K_{21} + e^{ - i \delta} K_{12} \right) 
\left( s^2_\psi e^{ - i h_{2} x } - c^2_\psi e^{ - i h_{1} x } \right) 
\biggr\}
\nonumber \\
&+& 
c^2_{\phi} c_\psi s_\psi 
\left[ \sin 2\psi \left( K_{11} - K_{22} \right) 
+ \cos 2\psi \left( e^{ i \delta} K_{21} + e^{ - i \delta} K_{12} \right) \right] 
\frac{ \Delta_{b} }{ h_{2} - h_{1} } \left\{ e^{ - i h_{2} x } - e^{ - i h_{1} x } \right\} 
\nonumber \\
&+& 
c_{\phi} s_{\phi} c_\psi 
\left\{ 
c_\psi \left( K_{31} + K_{13} \right) \Delta_{b} 
- s_\psi \left( e^{ i \delta} K_{23} + e^{ - i \delta} K_{32} \right) \Delta_{b} 
- 2 \widetilde{\epsilon} s_{\psi} \Delta_{ \text{ren} } \right\}
\frac{ 1 }{ h_{3} - h_{1} } \left\{ e^{ - i h_{3} x } - e^{ - i h_{1} x } \right\} 
\nonumber \\
&+& 
c_{\phi} s_{\phi} s_\psi 
\left\{ s_\psi \left( K_{31} + K_{13} \right) \Delta_{b} 
+ c_\psi \left( e^{ i \delta} K_{23} + e^{ - i \delta} K_{32} \right) \Delta_{b} 
+ 2 \widetilde{\epsilon} c_{\psi} \Delta_{ \text{ren} } \right\} 
\frac{ 1 }{ h_{3} - h_{2} } \left\{ e^{ - i h_{3} x } - e^{ - i h_{2} x } \right\}.
\nonumber 
\end{eqnarray}
\begin{eqnarray}
&& 
\widetilde{S}^{(1)}_{22} =
( - i \Delta_{b} x ) 
\biggl\{
c^2_\psi s^2_\psi K_{11} \left( e^{ - i h_{2} x } + e^{ - i h_{1} x } \right) 
+ K_{22} \left( c^4_\psi e^{ - i h_{2} x }  + s^4_\psi e^{ - i h_{1} x } \right) 
\nonumber \\
&+& 
c_\psi s_\psi \left( e^{ i \delta} K_{21} + e^{ - i \delta} K_{12} \right) 
\left( c^2_\psi e^{ - i h_{2} x } - s^2_\psi e^{ - i h_{1} x } \right) 
\biggr\}
\nonumber \\
&-& 
c_\psi s_\psi 
\left[
\sin 2\psi \left( K_{11} - K_{22} \right) 
+ \cos 2\psi \left( e^{ i \delta} K_{21} + e^{ - i \delta} K_{12} \right) 
\right]
\frac{ \Delta_{b} }{ h_{2} - h_{1} } \left\{ e^{ - i h_{2} x } - e^{ - i h_{1} x } \right\}. 
\nonumber 
\end{eqnarray}
\begin{eqnarray}
&& 
\widetilde{S}^{(1)}_{33} = 
( - i \Delta_{b} x ) 
\biggl\{
s^2_{\phi} K_{11} \left( s^4_\psi e^{ - i h_{2} x } + c^4_\psi e^{ - i h_{1} x } \right) 
+ s^2_{\phi} c^2_\psi s^2_\psi K_{22} \left( e^{ - i h_{2} x } + e^{ - i h_{1} x } \right)
\nonumber \\
&+& 
s^2_{\phi} c_\psi s_\psi \left( e^{ i \delta} K_{21} + e^{ - i \delta} K_{12} \right) 
\left( s^2_\psi e^{ - i h_{2} x } - c^2_\psi e^{ - i h_{1} x } \right) 
+ c^2_{\phi} K_{33} e^{ - i h_{3} x } 
\biggr\} 
\nonumber \\
&+& 
s^2_{\phi} c_\psi s_\psi 
\left[ \sin 2\psi \left( K_{11} - K_{22} \right) 
+ \cos 2\psi \left( e^{ i \delta} K_{21} + e^{ - i \delta} K_{12} \right) \right] 
\frac{ \Delta_{b} }{ h_{2} - h_{1} } \left\{ e^{ - i h_{2} x } - e^{ - i h_{1} x } \right\} 
\nonumber \\
&-& 
c_{\phi} s_{\phi} c_\psi 
\left\{ c_\psi \left( K_{13} + K_{31} \right) \Delta_{b} 
- s_\psi \left( e^{ i \delta} K_{23} + e^{ - i \delta} K_{32} \right) \Delta_{b} 
- 2 \widetilde{\epsilon} s_{\psi} \Delta_{ \text{ren} } \right\} 
\frac{ 1 }{ h_{3} - h_{1} } \left\{ e^{ - i h_{3} x } - e^{ - i h_{1} x } \right\} 
\nonumber \\
&-& 
c_{\phi} s_{\phi} s_\psi 
\left\{ s_\psi \left( K_{13} + K_{31} \right) \Delta_{b} 
+ c_\psi \left( e^{ i \delta} K_{23} + e^{ - i \delta} K_{32} \right) \Delta_{b} 
+ 2 \widetilde{\epsilon} c_{\psi} \Delta_{ \text{ren} } \right\} 
\frac{ 1 }{ h_{3} - h_{2} } \left\{ e^{ - i h_{3} x } - e^{ - i h_{2} x } \right\}.
\nonumber 
\end{eqnarray}
\begin{eqnarray}
&&
\widetilde{S}^{(1)}_{12} = 
e^{ i \delta} 
\biggl[
c_{\phi} c_\psi s_\psi 
( - i \Delta_{b} x ) 
\biggl\{
K_{11} \left( s^2_\psi e^{ - i h_{2} x } - c^2_\psi e^{ - i h_{1} x } \right) 
+ K_{22} \left( c^2_\psi e^{ - i h_{2} x } - s^2_\psi e^{ - i h_{1} x } \right) 
\nonumber \\
&+&
c_\psi s_\psi 
\left( e^{ i \delta} K_{21} + e^{ - i \delta} K_{12} \right) 
\left( e^{ - i h_{2} x } + e^{ - i h_{1} x } \right) 
\biggr\} 
\nonumber \\
&+& 
c_{\phi} \biggl\{ 
e^{ - i \delta} K_{12} 
+ c_\psi s_\psi 
\left[ \cos 2\psi \left( K_{11} - K_{22} \right) 
- \sin 2\psi \left( e^{ i \delta} K_{21} + e^{ - i \delta} K_{12} \right) \right] 
\biggr\} 
\frac{ \Delta_{b} }{ h_{2} - h_{1} } \left\{ e^{ - i h_{2} x } - e^{ - i h_{1} x } \right\} 
\nonumber \\
&-& 
s_{\phi} s_\psi 
\left\{ \left( c_\psi K_{31} - s_\psi e^{ - i \delta} K_{32} \right) 
\Delta_{b} - \widetilde{\epsilon} s_{\psi} \Delta_{ \text{ren} } \right\} 
\frac{ 1 }{ h_{3} - h_{1} } \left\{ e^{ - i h_{3} x } - e^{ - i h_{1} x } \right\} 
\nonumber \\
&+& 
s_{\phi} c_\psi 
\left\{ \left( s_\psi K_{31} + c_\psi e^{ - i \delta} K_{32} \right) 
\Delta_{b} + \widetilde{\epsilon} c_{\psi} \Delta_{ \text{ren} } \right\} 
\frac{ 1 }{ h_{3} - h_{2} } \left\{ e^{ - i h_{3} x } - e^{ - i h_{2} x } \right\} 
\biggr]. 
\nonumber 
\end{eqnarray}
\begin{eqnarray}
&&
\widetilde{S}^{(1)}_{21} = 
e^{ - i \delta} 
\biggl[
c_{\phi} c_\psi s_\psi ( - i \Delta_{b} x ) 
\biggl\{
K_{11} \left( s^2_\psi e^{ - i h_{2} x } - c^2_\psi e^{ - i h_{1} x } \right) 
+ K_{22} \left( c^2_\psi e^{ - i h_{2} x } - s^2_\psi e^{ - i h_{1} x } \right) 
\nonumber \\
&+&
c_\psi s_\psi 
\left( e^{ i \delta} K_{21} + e^{ - i \delta} K_{12} \right) 
\left( e^{ - i h_{2} x } + e^{ - i h_{1} x } \right) 
\biggr\} 
\nonumber \\
&+& 
c_{\phi} \biggl\{ 
e^{ i \delta} K_{21} 
+ c_\psi s_\psi  
\left[ \cos 2\psi \left( K_{11} - K_{22} \right) 
- \sin 2\psi \left( e^{ i \delta} K_{21} + e^{ - i \delta} K_{12} \right) \right] 
\biggr\} 
\frac{ \Delta_{b} }{ h_{2} - h_{1} } \left\{ e^{ - i h_{2} x } - e^{ - i h_{1} x } \right\} 
\nonumber \\
&-& 
s_{\phi} s_\psi 
\left\{ \left( c_\psi K_{13} - s_\psi e^{ i \delta} K_{23} \right) 
\Delta_{b} - \widetilde{\epsilon} s_{\psi} \Delta_{ \text{ren} } \right\} 
\frac{ 1 }{ h_{3} - h_{1} } \left\{ e^{ - i h_{3} x } - e^{ - i h_{1} x } \right\} 
\nonumber \\
&+& 
s_{\phi} c_\psi 
\left\{ \left( s_\psi K_{13} + c_\psi e^{ i \delta} K_{23} \right) 
\Delta_{b} + \widetilde{\epsilon} c_{\psi} \Delta_{ \text{ren} } \right\} 
\frac{ 1 }{ h_{3} - h_{2} } \left\{ e^{ - i h_{3} x } - e^{ - i h_{2} x } \right\} 
\biggr]. 
\nonumber 
\end{eqnarray}
\begin{eqnarray}
&& 
\widetilde{S}^{(1)}_{13} = 
- ( - i \Delta_{b} x ) c_{\phi} s_{\phi} 
\biggl\{
K_{11} \left( s^4_\psi e^{ - i h_{2} x } + c^4_\psi e^{ - i h_{1} x } \right) 
+ c^2_\psi s^2_\psi K_{22} \left( e^{ - i h_{2} x } + e^{ - i h_{1} x } \right) 
- K_{33} e^{ - i h_{3} x } 
\nonumber \\
&+& 
c_\psi s_\psi \left( e^{ i \delta} K_{21} + e^{ - i \delta} K_{12} \right) 
\left( s^2_\psi e^{ - i h_{2} x } - c^2_\psi e^{ - i h_{1} x } \right) 
\biggr\}
\nonumber \\
&-& 
c_{\phi} s_{\phi} 
c_\psi s_\psi 
\left[ \sin 2\psi \left( K_{11} - K_{22} \right) 
+ \cos 2\psi \left( e^{ i \delta} K_{21} + e^{ - i \delta} K_{12} \right) \right] 
\frac{ \Delta_{b} }{ h_{2} - h_{1} } \left\{ e^{ - i h_{2} x } - e^{ - i h_{1} x } \right\} 
\nonumber \\
&& 
\hspace{-15mm}
+ \left\{ 
c^2_{\psi} \left( c^2_{\phi} K_{13} - s^2_{\phi} K_{31} \right) \Delta_{b} 
- c_\psi s_\psi \left( c^2_{\phi} e^{ i \delta} K_{23} - s^2_{\phi} e^{ - i \delta} K_{32} \right) \Delta_{b} 
- \cos 2\phi c_\psi \widetilde{\epsilon} s_{\psi} \Delta_{ \text{ren} } 
\right\} 
\frac{ 1 }{ h_{3} - h_{1} } \left\{ e^{ - i h_{3} x } - e^{ - i h_{1} x } \right\} 
\nonumber \\
&& 
\hspace{-15mm}
+ \left\{ s^2_{\psi} \left( c^2_{\phi} K_{13} - s^2_{\phi} K_{31} \right) \Delta_{b} 
+ c_\psi s_\psi \left( c^2_{\phi} e^{ i \delta} K_{23} - s^2_{\phi} e^{ - i \delta} K_{32} \right) \Delta_{b} 
+ \cos 2\phi s_\psi \widetilde{\epsilon} c_{\psi} \Delta_{ \text{ren} } 
\right\} 
\frac{ 1 }{ h_{3} - h_{2} } \left\{ e^{ - i h_{3} x } - e^{ - i h_{2} x } \right\} 
\nonumber 
\end{eqnarray} 
\begin{eqnarray}
&&
\widetilde{S}^{(1)}_{31} = 
- ( - i \Delta_{b} x ) c_{\phi} s_{\phi} 
\biggl\{
K_{11} \left( s^4_\psi e^{ - i h_{2} x } + c^4_\psi e^{ - i h_{1} x } \right) 
+ c^2_\psi s^2_\psi K_{22} \left( e^{ - i h_{2} x } + e^{ - i h_{1} x } \right) 
- K_{33} e^{ - i h_{3} x } 
\nonumber \\
&+& 
c_\psi s_\psi \left( e^{ i \delta} K_{21} + e^{ - i \delta} K_{12} \right) 
\left( s^2_\psi e^{ - i h_{2} x } - c^2_\psi e^{ - i h_{1} x } \right) 
\biggr\}
\nonumber \\
&-& 
c_{\phi} s_{\phi} 
c_\psi s_\psi 
\left[ \sin 2\psi \left( K_{11} - K_{22} \right) 
+ \cos 2\psi \left( e^{ i \delta} K_{21} + e^{ - i \delta} K_{12} \right) \right] 
\frac{ \Delta_{b} }{ h_{2} - h_{1} } \left\{ e^{ - i h_{2} x } - e^{ - i h_{1} x } \right\} 
\nonumber \\
&& 
\hspace{-15mm}
+ \left\{ 
c^2_\psi \left( c^2_{\phi} K_{31} - s^2_{\phi} K_{13} \right) \Delta_{b} 
+ c_\psi s_\psi \left( s^2_{\phi} e^{ i \delta} K_{23} - c^2_{\phi} e^{ - i \delta} K_{32} \right) \Delta_{b} 
- \cos 2\phi c_\psi \widetilde{\epsilon} s_{\psi} \Delta_{ \text{ren} } 
\right\} 
\frac{ 1 }{ h_{3} - h_{1} } \left\{ e^{ - i h_{3} x } - e^{ - i h_{1} x } \right\} 
\nonumber \\
&& 
\hspace{-15mm}
+ \left\{ 
s^2_\psi \left( c^2_{\phi} K_{31} - s^2_{\phi} K_{13} \right) \Delta_{b} 
- c_\psi s_\psi \left( s^2_{\phi} e^{ i \delta} K_{23} - c^2_{\phi} e^{ - i \delta} K_{32} \right) \Delta_{b} 
+ \cos 2\phi s_\psi \widetilde{\epsilon} c_{\psi} \Delta_{ \text{ren} } 
\right\} 
\frac{ 1 }{ h_{3} - h_{2} } \left\{ e^{ - i h_{3} x } - e^{ - i h_{2} x } \right\}. 
\nonumber 
\end{eqnarray}
\begin{eqnarray}
&& 
\widetilde{S}^{(1)}_{23} = 
e^{ - i \delta} 
\biggl[
- ( - i \Delta_{b} x ) s_{\phi} c_\psi s_\psi 
\biggl\{
K_{11} \left( s^2_\psi e^{ - i h_{2} x } - c^2_\psi e^{ - i h_{1} x } \right) 
+ K_{22} \left( c^2_\psi e^{ - i h_{2} x } - s^2_\psi e^{ - i h_{1} x } \right) 
\nonumber \\
&+&
c_\psi s_\psi 
\left( e^{ i \delta} K_{21} + e^{ - i \delta} K_{12} \right) 
\left( e^{ - i h_{2} x } + e^{ - i h_{1} x } \right) 
\biggr\} 
\nonumber \\
&-& 
s_{\phi} \biggl\{ 
e^{ i \delta} K_{21} 
+ c_\psi s_\psi 
\left[ \cos 2\psi \left( K_{11} - K_{22} \right) 
- \sin 2\psi \left( e^{ i \delta} K_{21} + e^{ - i \delta} K_{12} \right) \right] 
\biggr\} 
\frac{ \Delta_{b} }{ h_{2} - h_{1} } \left\{ e^{ - i h_{2} x } - e^{ - i h_{1} x } \right\} 
\nonumber \\
&-&
c_{\phi} s_\psi 
\left\{ \left( c_\psi K_{13} - s_\psi e^{ i \delta} K_{23} \right) 
\Delta_{b} - \widetilde{\epsilon} s_{\psi} \Delta_{ \text{ren} } \right\} 
\frac{ 1 }{ h_{3} - h_{1} } \left\{ e^{ - i h_{3} x } - e^{ - i h_{1} x } \right\} 
\nonumber \\
&+& 
c_{\phi} c_\psi 
\left\{ \left( s_\psi K_{13} + c_\psi e^{ i \delta} K_{23} \right) 
\Delta_{b} + \widetilde{\epsilon} c_{\psi} \Delta_{ \text{ren} } \right\} 
\frac{ 1 }{ h_{3} - h_{2} } \left\{ e^{ - i h_{3} x } - e^{ - i h_{2} x } \right\} 
\biggr].
\nonumber 
\end{eqnarray}
\begin{eqnarray}
&&
\widetilde{S}^{(1)}_{32} = 
e^{ i \delta} 
\biggl[
- ( - i \Delta_{b} x ) s_{\phi} c_\psi s_\psi 
\biggl\{
K_{11} \left( s^2_\psi e^{ - i h_{2} x } - c^2_\psi e^{ - i h_{1} x } \right) 
+ K_{22} \left( c^2_\psi e^{ - i h_{2} x } - s^2_\psi e^{ - i h_{1} x } \right) 
\nonumber \\
&+&
c_\psi s_\psi 
\left( e^{ i \delta} K_{21} + e^{ - i \delta} K_{12} \right) 
\left( e^{ - i h_{2} x } + e^{ - i h_{1} x } \right) 
\biggr\} 
\nonumber \\
&-& 
s_{\phi} \biggl\{ 
e^{ - i \delta} K_{12} 
+ c_\psi s_\psi 
\left[ \cos 2\psi \left( K_{11} - K_{22} \right) 
- \sin 2\psi \left( e^{ i \delta} K_{21} + e^{ - i \delta} K_{12} \right) \right] 
\biggr\} 
\frac{ \Delta_{b} }{ h_{2} - h_{1} } \left\{ e^{ - i h_{2} x } - e^{ - i h_{1} x } \right\} 
\nonumber \\
&-&
c_{\phi} s_\psi 
\left\{ \left( c_\psi K_{31} - s_\psi e^{ - i \delta} K_{32} \right) 
\Delta_{b} - \widetilde{\epsilon} s_{\psi} \Delta_{ \text{ren} } \right\} 
\frac{ 1 }{ h_{3} - h_{1} } \left\{ e^{ - i h_{3} x } - e^{ - i h_{1} x } \right\} 
\nonumber \\
&+& 
c_{\phi} c_\psi 
\left\{ \left( s_\psi K_{31} + c_\psi e^{ - i \delta} K_{32} \right) 
\Delta_{b} + \widetilde{\epsilon} c_{\psi} \Delta_{ \text{ren} } \right\} 
\frac{ 1 }{ h_{3} - h_{2} } \left\{ e^{ - i h_{3} x } - e^{ - i h_{2} x } \right\} 
\biggr].
\nonumber 
\end{eqnarray}
$\widetilde{\epsilon}$ is defined in eq.~\eqref{widetilde-epsilon-def}. As in the zeroth-order $\widetilde{S}^{(0)}$ matrix, the generalized T invariance holds as well.

\section{First-order unitary evolution part $P(\nu_{\mu} \rightarrow \nu_{e})_{ \text{ EV } }^{(1)}$ in the atmospheric resonance region} 
\label{sec:P-mue-EV-1st-atm}

The probability $P(\nu_{\mu} \rightarrow \nu_{e})_{ \text{EV} }^{(1)}$ in eq.~\eqref{P-mue-EV-ATM} with use of the $\widetilde{\alpha}$ parameters is given by 
\begin{eqnarray} 
&& 
P(\nu_{\mu} \rightarrow \nu_{e})_{ \text{EV} }^{(1)} 
= 2 \mbox{Re} \left[
\left( S^{(0)}_{e \mu} \right)^* 
\left( S^{(1)}_{ \text{EV} } \right)_{e \mu} 
\right] 
\nonumber \\
&=& 
s^2_{23} \sin^2 2\phi 
\biggl[ 
\cos 2\phi \biggl\{ 
- \widetilde{\alpha}_{ee} 
\left( 1 - \frac{ \Delta_{a} }{ \Delta_{b} } \right) 
+  \left[ s_{23}^2 \widetilde{\alpha}_{\mu \mu} + c_{23}^2 \widetilde{\alpha}_{\tau \tau} 
+ c_{23} s_{23} \mbox{Re} \left( \widetilde{\alpha}_{\tau \mu} \right) \right] 
\biggr\} 
\nonumber \\
&+& 
\sin 2\phi 
\left[ s_{23} \mbox{Re} \left( \widetilde{\alpha}_{\mu e} \right) 
+ c_{23} \mbox{Re} \left( \widetilde{\alpha}_{\tau e} \right) 
\right] 
\biggr] 
( \Delta_{b} x ) \sin ( h_{3} - h_{1} ) x 
%
\nonumber \\
&+& 
\sin 2\theta_{23} \sin 2\phi 
\biggl\{ 
c^2_{\phi} \left[ c_{23} \mbox{Re} \left( \widetilde{\alpha}_{\mu e} \right) 
- s_{23} \mbox{Re} \left( \widetilde{\alpha}_{\tau e} \right) \right] 
- c_{\phi} s_{\phi} 
\left[ \sin 2\theta_{23} ( \widetilde{\alpha}_{\mu \mu} - \widetilde{\alpha}_{\tau \tau} ) 
+ \cos 2\theta_{23} \mbox{Re} \left( \widetilde{\alpha}_{\tau \mu} \right) \right] 
\biggr\} 
\nonumber \\
&\times& 
\frac{ \Delta_{b} }{ h_{2} - h_{1} } 
\left\{ - \sin^2 \frac{ ( h_{3} - h_{2} ) x }{2} 
+ \sin^2 \frac{ ( h_{3} - h_{1} ) x }{2} 
+ \sin^2 \frac{ ( h_{2} - h_{1} ) x }{2} \right\} 
%
\nonumber \\
&+& 
\sin 2\theta_{23} \sin 2\phi 
\biggl\{ 
s^2_{\phi} 
\left[ c_{23} \mbox{Re} \left( \widetilde{\alpha}_{\mu e} \right) 
- s_{23} \mbox{Re} \left( \widetilde{\alpha}_{\tau e} \right) \right] 
+ c_{\phi} s_{\phi}
\left[ \sin 2\theta_{23} ( \widetilde{\alpha}_{\mu \mu} - \widetilde{\alpha}_{\tau \tau} ) 
+ \cos 2\theta_{23} \mbox{Re} \left( \widetilde{\alpha}_{\tau \mu} \right) \right] 
\biggr\}
\nonumber \\
&\times& 
\frac{ \Delta_{b} }{ h_{3} - h_{2} } 
\left\{ 
\sin^2 \frac{ ( h_{3} - h_{2} ) x }{2} 
+ \sin^2 \frac{ ( h_{3} - h_{1} ) x }{2} 
- \sin^2 \frac{ ( h_{2} - h_{1} ) x }{2} 
\right\} 
%
\nonumber \\
&+& 
4 s^2_{23} \cos 2\phi \sin 2\phi 
\biggl\{
\sin 2\phi 
\left[ 
\widetilde{\alpha}_{ee} \left( 1 - \frac{ \Delta_{a} }{ \Delta_{b} } \right) 
- \left( s_{23}^2 \widetilde{\alpha}_{\mu \mu} + c_{23}^2 \widetilde{\alpha}_{\tau \tau} \right) 
- c_{23} s_{23} 
\mbox{Re} \left( \widetilde{\alpha}_{\tau \mu} \right) 
\right] 
\nonumber \\
&+& 
\cos 2\phi 
\left[ s_{23} \mbox{Re} \left( \widetilde{\alpha}_{\mu e} \right) + c_{23} \mbox{Re} \left( \widetilde{\alpha}_{\tau e} \right) \right] 
\biggr\} 
\frac{ \Delta_{b} }{ h_{3} - h_{1} } 
\sin^2 \frac{ ( h_{3} - h_{1} ) x }{2} 
%
\nonumber \\
&+& 
2 \sin 2\theta_{23} \sin 2\phi 
\biggl\{ 
c^2_{\phi} 
\left[ c_{23} \mbox{Im} \left( \widetilde{\alpha}_{\mu e} \right) 
- s_{23} \mbox{Im} \left( \widetilde{\alpha}_{\tau e} \right) \right] 
+ c_{\phi} s_{\phi} \mbox{Im} \left( \widetilde{\alpha}_{\tau \mu} \right)  
\biggr\} 
\nonumber \\
&\times& 
\frac{ \Delta_{b} }{ h_{2} - h_{1} } 
\sin \frac{ ( h_{3} - h_{1} ) x }{2} \sin \frac{ ( h_{2} - h_{1} ) x }{2} \sin \frac{ ( h_{3} - h_{2} ) x }{2} 
%
\nonumber \\
&-& 
2 \sin 2\theta_{23} \sin 2\phi 
\biggl\{ 
s^2_{\phi} 
\left[ c_{23} \mbox{Im} \left( \widetilde{\alpha}_{\mu e} \right) 
- s_{23} \mbox{Im} \left( \widetilde{\alpha}_{\tau e} \right) \right] 
- c_{\phi} s_{\phi} \mbox{Im} \left( \widetilde{\alpha}_{\tau \mu} \right) 
\biggr\}
\nonumber \\
&\times& 
\frac{ \Delta_{b} }{ h_{3} - h_{2} } 
\sin \frac{ ( h_{3} - h_{1} ) x }{2} \sin \frac{ ( h_{2} - h_{1} ) x }{2} \sin \frac{ ( h_{3} - h_{2} ) x }{2}, 
%
\label{P-mue-EV-1st-alpha}
\end{eqnarray}
which reproduces eq.~(49) in ref.~\cite{Martinez-Soler:2018lcy}.

\section{Flavor basis $S$ matrix in the $\nu_{\mu} \rightarrow \nu_{\tau}$ channel} 
\label{sec:S-taumu}

The flavor basis $S$ matrix in the $\nu_{\mu} \rightarrow \nu_{\tau}$ channel is given in the first order in the $\widetilde{\alpha}$ parameters (ignoring the first-order $\nu$SM part) as 
\begin{eqnarray}
&& 
S_{\tau \mu}^{(1)} 
= 
c^2_{23} \widetilde{S}^{(1)}_{32} - s^2_{23} \widetilde{S}^{(1)}_{23} 
+ c_{23} s_{23} \left( \widetilde{S}^{(1)}_{33} - \widetilde{S}^{(1)}_{22} \right) 
\nonumber \\
&=& 
- s_{\phi} c_\psi s_\psi 
\left( c^2_{23} e^{ i \delta} - s^2_{23} e^{ - i \delta} \right) 
( - i \Delta_{b} x ) 
\biggl\{
K_{11} \left( s^2_\psi e^{ - i h_{2} x } - c^2_\psi e^{ - i h_{1} x } \right) 
+ K_{22} \left( c^2_\psi e^{ - i h_{2} x } - s^2_\psi e^{ - i h_{1} x } \right) 
\nonumber \\
&+&
c_\psi s_\psi 
\left( e^{ i \delta} K_{21} + e^{ - i \delta} K_{12} \right) 
\left( e^{ - i h_{2} x } + e^{ - i h_{1} x } \right) 
\biggr\} 
\nonumber \\
&+& 
c_{23} s_{23} ( - i \Delta_{b} x ) 
\biggl\{ 
\left[ 
s^2_\psi K_{11} + c^2_\psi K_{22} 
+ c_\psi s_\psi \left( e^{ i \delta} K_{21} + e^{ - i \delta} K_{12} \right) 
\right] 
( s^2_{\phi} s^2_\psi - c^2_\psi ) e^{ - i h_{2} x } 
\nonumber \\
&+& 
\left[ 
c^2_\psi K_{11} + s^2_\psi K_{22} 
- c_\psi s_\psi \left( e^{ i \delta} K_{21} + e^{ - i \delta} K_{12} \right) 
\right] 
( s^2_{\phi} c^2_\psi - s^2_\psi ) e^{ - i h_{1} x } 
+ c^2_{\phi} K_{33} e^{ - i h_{3} x } 
\biggr\} 
\nonumber \\
&& 
\hspace{-12mm}
- s_{\phi} 
\biggl[ 
\cos 2\theta_{23}
\left\{
\cos \delta \mbox{Re} \left( e^{ - i \delta} K_{12} \right) 
- \sin \delta \mbox{Im} \left( e^{ - i \delta} K_{12} \right) 
\right\} 
+ i \left\{ \sin \delta \mbox{Re} \left( e^{ - i \delta} K_{12} \right) 
+ \cos \delta \mbox{Im} \left( e^{ - i \delta} K_{12} \right) \right\} 
\biggr]
\nonumber \\ 
&\times& 
\frac{ \Delta_{b} }{ h_{2} - h_{1} } \left\{ e^{ - i h_{2} x } - e^{ - i h_{1} x } \right\} 
\nonumber \\
&& 
\hspace{-12mm}
- \left( c^2_{23} e^{ i \delta} - s^2_{23} e^{ - i \delta} \right) 
s_{\phi} c_\psi s_\psi 
\left[ \cos 2\psi \left( K_{11} - K_{22} \right) 
- \sin 2\psi \left( e^{ i \delta} K_{21} + e^{ - i \delta} K_{12} \right) \right] 
\frac{ \Delta_{b} }{ h_{2} - h_{1} } \left\{ e^{ - i h_{2} x } - e^{ - i h_{1} x } \right\} 
\nonumber \\
&+& 
c_{23} s_{23} \sin 2\psi 
\left[ \sin 2\psi \left( K_{11} - K_{22} \right) 
+ \cos 2\psi \left( e^{ i \delta} K_{21} + e^{ - i \delta} K_{12} \right) \right] 
\frac{ \Delta_{b} }{ h_{2} - h_{1} } \left\{ e^{ - i h_{2} x } - e^{ - i h_{1} x } \right\} 
\nonumber \\
&-& 
c_{\phi} s_\psi 
\biggl[
\cos 2\theta_{23}
\left\{
\cos \delta \mbox{Re} \left( c_\psi K_{31} - s_\psi e^{ - i \delta} K_{32} \right) 
- \sin \delta \mbox{Im} \left( c_\psi K_{31} - s_\psi e^{ - i \delta} K_{32} \right) 
\right\} 
\nonumber \\
&+& 
i \left\{ \sin \delta \mbox{Re} \left( c_\psi K_{31} - s_\psi e^{ - i \delta} K_{32} \right) 
+ \cos \delta \mbox{Im} \left( c_\psi K_{31} - s_\psi e^{ - i \delta} K_{32} \right) \right\} 
\biggr] 
\frac{ \Delta_{b} }{ h_{3} - h_{1} } \left\{ e^{ - i h_{3} x } - e^{ - i h_{1} x } \right\} 
\nonumber \\
&+& 
c_{\phi} c_\psi 
\biggl[
\cos 2\theta_{23}
\left\{
\cos \delta \mbox{Re} \left( s_\psi K_{31} + c_\psi e^{ - i \delta} K_{32} \right) 
- \sin \delta \mbox{Im} \left( s_\psi K_{31} + c_\psi e^{ - i \delta} K_{32} \right) 
\right\} 
\nonumber \\
&+& 
i \left\{ \sin \delta \mbox{Re} \left( s_\psi K_{31} + c_\psi e^{ - i \delta} K_{32} \right) 
+ \cos \delta \mbox{Im} \left( s_\psi K_{31} + c_\psi e^{ - i \delta} K_{32} \right) \right\} 
\biggr] 
\frac{ \Delta_{b} }{ h_{3} - h_{2} } \left\{ e^{ - i h_{3} x } - e^{ - i h_{2} x } \right\} 
\nonumber \\
&-& 
c_{23} s_{23} c_{\phi} s_{\phi} c_\psi 
\left[ c_\psi \left( K_{13} + K_{31} \right)  
- s_\psi \left( e^{ i \delta} K_{23} + e^{ - i \delta} K_{32} \right) \right] 
\frac{ \Delta_{b} }{ h_{3} - h_{1} } \left\{ e^{ - i h_{3} x } - e^{ - i h_{1} x } \right\} 
\nonumber \\
&-& 
c_{23} s_{23} c_{\phi} s_{\phi} s_\psi 
\left[ s_\psi \left( K_{13} + K_{31} \right) 
+ c_\psi \left( e^{ i \delta} K_{23} + e^{ - i \delta} K_{32} \right) \right] 
\frac{ \Delta_{b} }{ h_{3} - h_{2} } \left\{ e^{ - i h_{3} x } - e^{ - i h_{2} x } \right\}. 
\label{flavor-S-taumu-1st}
\end{eqnarray}

\end{document}